\documentclass{article}   

\newcommand{\dsty}{\displaystyle}
\newcommand{\e}{{\mathrm e}}

\renewcommand{\i}{{\mathrm i}}

\renewcommand{\d}{{\mathrm d}}
\def\bea{\begin{eqnarray}}
\def\eea{\end{eqnarray}}
\def\tfrac#1#2{{\textstyle {#1 \over #2}}}
\def\tsum{\mathop{\textstyle \sum }}
\def\tprod{\mathop{\textstyle \prod }}
\def\binom#1#2{{#1 \choose #2}}
\def\QATOPD#1#2#3#4{{#3 \atopwithdelims#1#2 #4}}

\title{Integrals involving triplets of Jacobi and Gegenbauer polynomials 
and some 3$j$-symbols of SO($n$), SU($n$) and Sp(4)}
\author{Sigitas Ali\v sauskas }
\date{Institute of Theoretical Physics and Astronomy \\ 
of Vilnius University, \\
A. Go\v stauto 12, LT-01108 Vilnius, Lithuania}
\makeatletter
\newcounter{parentequation}
\@ifundefined{ignorespacesafterend}{%
  \def\ignorespacesafterend{\global\@ignoretrue}%
}{}
\newenvironment{subequations}{%
  \refstepcounter{equation}%
  \protected@edef\theparentequation{\theequation}%
  \setcounter{parentequation}{\value{equation}}%
  \setcounter{equation}{0}%
  \def\theequation{\theparentequation\alph{equation}}%
  \ignorespaces
}{%
  \setcounter{equation}{\value{parentequation}}%
  \ignorespacesafterend
}
\makeatother

\begin{document}           

\maketitle                 

The coupling coefficients (3$j$-symbols) for the symmetric (most 
degenerate) irreducible representations of the orthogonal groups SO($n$) 
in a canonical basis [with SO($n$) restricted to SO($n-1$)] and different 
semicanonical (tree) bases [with SO($n$) restricted to SO$(n^{\prime })
\times $SO$(n^{\prime \prime })$, $n^{\prime }+n^{\prime \prime }=n$] 
are expressed in terms of the integrals involving triplets of the 
Gegenbauer and the Jacobi polynomials. The derived usual 
triple-hypergeometric series (which do not reveal the apparent triangle 
conditions of the 3$j$-symbols) are rearranged directly [without using 
their relation with the semistretched isofactors of the second kind for 
the complementary chain Sp(4)$\supset $SU(2)$\times $SU(2)] into formulas 
with more rich limits for summation intervals and obvious triangle 
conditions. The isofactors for the class-one representations of the 
orthogonal groups and for the class-two representations of the unitary 
groups (and, of course, the related integrals) turn into the double sums 
in the cases of the canonical SO$(n)\!\supset $SO$(n-1)$ or 
U$(n)\!\supset $U$(n-1)$ and semicanonical SO$(n)\!\supset $SO$(n-2)\times $%
SO(2) chains, as well as into the $_4F_3(1)$ series under more specific 
conditions. Expressions for the most general isofactors of SO($n$) for 
coupling of the two symmetric irreps in the canonical basis are also 
derived.

\section{Introduction}

The Clebsch--Gordan (coupling) coefficients and $3j$-symbols (or the 
Wigner coefficients) of the orthogonal groups SO($n$), together with their 
isoscalar factors (isofactors), maintain great importance in many fields 
of theoretical physics such as atomic, nuclear and statistical physics. 
The representation functions in terms of the Gegenbauer (ultraspherical) 
polynomials are well known for the symmetric (also called most degenerate 
or class-one) irreducible representations (irreps) of SO$(n)$ in the 
spherical coordinates (Vilenkin \cite{Vi65a}) on the unit sphere 
$S_{n-1}$. In particular, the explicit Clebsch--Gordan (CG) coefficients 
and isofactors of SO($n$) in the canonical basis for all three symmetric 
irreps were considered by Gavrilik \cite{Ga73}, Kildyushov and Kuznetsov 
\cite{KK73} (see also \cite{KMSS92}) and Junker \cite{Ju93}, using the 
direct \cite{Ga73,Ju93} or rather complicated indirect \cite{KK73,KMSS92} 
integration procedures.

Norvai\v sas and Ali\v sauskas \cite{NAl74a} also derived triple-sum 
expressions for related isofactors of SO($n$) in the case of 
the canonical (labelled by the chain of groups SO$(n)\!\supset $SO($n-1$)) 
and semicanonical bases (labelled by irreps $l,l^{\prime },
l^{\prime \prime }$ of the group chains SO$(n)\!\supset $SO$(n^{\prime 
})\times $SO$(n^{\prime \prime })$, $n^{\prime }+n^{\prime \prime }=n$, 
in the polyspherical, or the tree type, coordinates 
\cite{Vi65a,KMSS92,Vi65b}), exploiting the transition matrices 
\cite{AlV72} (also cf.\ \cite{K97}) between the bases, labelled by the 
unitary and orthogonal subgroups in the symmetrical irreducible spaces of 
the U$(n)$ group. They have observed \cite{NAl74a,NAl74b} that isofactors 
for the group chain SO$(n)\!\supset $SO$(n^{\prime })\times $SO$(n^{%
\prime \prime })$ for the coupling of the states of symmetric irreps 
$l_1,l_2$ into the states of more general irreps $[L_1L_2]$ are the 
analytical continuation of the isofactors for the chain 
Sp(4)$\supset $SU(2)$\times $SU(2),
\bea
& {\left[ \begin{array}{ccc}
l_1 & l_2 & [L_1L_2] \\ 
l_1^{\prime },l_1^{\prime \prime } & l_2^{\prime },l_2^{\prime
\prime } & \gamma [L_1^{\prime }L_2^{\prime }][L_1^{\prime \prime
}L_2^{\prime \prime }]
\end{array} \right] _{(n:n^{\prime }n^{\prime \prime })}=(-1)^{\phi }} & 
\nonumber \\
& {\times \left[ \begin{array}{ccc}
\!\left\langle \frac{-2L_2^{\prime }-n^{\prime }}{4}\frac{-2L_1^{\prime
}-n^{\prime }}{4}\right\rangle \! & \!\left\langle \frac{-2L_2^{\prime 
\prime }-n^{\prime \prime }}{4}\frac{-2L_1^{\prime \prime }-n^{\prime 
\prime }}{4}\right\rangle \! & \!\left\langle \frac{-2L_2-n}{4}\frac{-2L_1
-n}{4}\right\rangle ^{\!\gamma }\! \\ 
\frac{-2l_1^{\prime }-n^{\prime }}{4},\frac{-2l_2^{\prime }-n^{\prime %
}}{4} & \frac{-2l_1^{\prime \prime }-n^{\prime \prime }}{4},\frac{%
-2l_2^{\prime \prime }-n^{\prime \prime }}{4} & \frac{-2l_1-n}{4},
\frac{-2l_2-n}{4}
\end{array} \right] ,} \label{anct}
\eea 
i.e.\ they coincide, up to phase factor $(-1)^{\phi }$, with the 
isofactors for the non-compact complementary group \cite{MQ70,MQ71,Q73} 
chain Sp(4,$R$)$\supset $Sp(2,$R$)$\times $Sp(2,$R$) in the case for the 
discrete series of irreps. Particularly, in a special multiplicity-free 
case (for $L_2=L_2^{\prime }=L_2^{\prime \prime }=0$, when the 
multiplicity label $\gamma $ is absent), the isofactors of 
SO$(n)\!\supset $SO($n-1$) correspond to the semistretched isofactors 
of the second kind \cite{AlJ71} of Sp(4)$\supset $SU(2)$\times $SU(2) 
(see also \cite{Al83,Al87}). 

However, neither expressions derived by means of direct integration
\cite{Ga73,Ju93}, nor the expressions derived by the re-expansion of the 
states of the group chains \cite{NAl74a,NAl74b} reveal the apparent 
triangle conditions of the 3j-symbols in these triple-sum series. Only 
the substitution group technique of the Sp(4) or SO(5) group \cite{AlJ69},
used together with an analytical continuation procedure, enabled the 
transformations of the initial triple-sum expressions of \cite{NAl74a} 
into other forms \cite{NAl74a,NAl74b,Al83,Al02a}, more convenient in 
the cases close to the stretched ones (e.g.\ for small values of shift 
$l_1+l_2-l_3$, where $l_3=L_1$) and turning into the double sums for 
the canonical basis, the SO$(n)\!\supset $SO$(n-2)\times $SO(2) 
chain and other cases with specified parameters 
$l_1^{\prime \prime }+l_2^{\prime \prime }-l_3^{\prime \prime }=0$ (where 
$l_3^{\prime \prime }=L_1^{\prime \prime }$). More specified isofactors 
of SO$(n)\!\supset $SO($n-1$) \cite{Al87} are related to $6j$ 
coefficients of SU(2) (with some parameters being quartervalued, i.e.\ 
multiple of 1/4, in the case when $n$ is odd).

Unfortunately, the empirical phase choices of isofactors in early 
publications \cite{NAl74a,NAl74b,Al83,Al87} were not correlated with the 
basis states (cf.\ \cite{Vi65a,KMSS92,IPSW01,MS01}) in terms of the 
Gegenbauer and the Jacobi polynomials and some aspects of the isofactor 
symmetry problem were left untouched (including the sign change for 
irreps $m$ of the SO(2) subgroups not revealed also in 
\cite{Vi65a,KMSS92,IPSW01} for the states of SO(3)$\supset $SO(2) and 
SO$(n)\!\supset $SO$(n-2)\times $SO(2)). Besides some indefiniteness 
of the double factorials in the numerator or denominator must be 
eliminated.

Recently the author has returned to the problem. The unambiguous proof of 
the most preferable and consistent expressions for the $3j$-symbols of 
the orthogonal SO($n$) and unitary U($n$) groups for decomposition of the 
factorized ultraspherical and polyspherical harmonics (i.e.\ for the 
coupling of three most degenerate irreps into scalar representation in 
the cases of the canonical and semicanonical bases) was reconsidered in 
\cite{Al02a} (cf.\ also \cite{Ga73,KK73,KMSS92,Ju93,NAl74a,Al87}), 
together with a comprehensive review of some adjusted previous results 
\cite{Ju93,NAl74a,AlJ71,Al83}, taking into account that some references 
\cite{Ga73,KK73,KMSS92,NAl74a,Al83} may be not easily accessible nor 
free from misprints. 

However, the main goal of \cite{Al02a} and this paper is a strict 
{\it ab initio} rearrangement of the most symmetric (although banal) 
finite triple-sum series of the hypergeometric-type in the expressions 
of the definite integrals involving triplets of the multiplied Gegenbauer 
and the Jacobi polynomials into less symmetric but more convenient triple 
(\ref{iJpd}), double (\ref{iGpco}), or single (\ref{iGpr}) sum series 
with the summation intervals depending on the triangular conditions of 
the corresponding $3j$-symbols. The related triple-hypergeometric series, 
appearing in the expressions \cite{AlJ71} for the semistretched isoscalar 
factors of the second kind of the chain Sp(4)$\supset $SU(2)$\times $SU(2), 
were considered in section 2 of \cite{Al02a}, together with their 
{\it ab initio} rearrangement using the different expressions \cite{Al00} 
for the stretched $9j$ coefficients of SU(2). (These triple sum series may 
be treated as extensions of the double-hypergeometric series of Kamp\'{e} 
de F\'{e}riet \cite{K-F21,AK-F26} type, e.g.\ considered by Lievens and 
Van der Jeugt \cite{LV-J01}.)

The well-known special integral involving triplet of the Jacobi 
polynomials $P_k^{(\alpha ,\beta )}(x)$ \cite{BE53,AS65,AAR99} in terms 
of the Clebsch--Gordan coefficients of SU(2)
\bea
& \dsty \frac 12\int\limits_{-1}^1\d x\left( \tfrac{1+x}{2}\right)
^{(\beta _1+\beta _2+\beta _3)/2}\left( \tfrac{1-x}{2}\right) 
^{(\alpha _1+\alpha _2+\alpha _3)/2}\prod_{a=1}^{3}P_{k_a}
^{(\alpha _a,\beta _a)}(x) &  \nonumber \\
& =\dsty \left[ \frac{1}{2l_3+1}\prod_{a=1}^{3}\frac{(k_a+\alpha _a)!
(k_a+\beta _a)!}{k_a!(k_a+\alpha _a+\beta _a)!}\right] ^{1/2} 
C_{m_1 m_2 m_3}^{l_1 l_2 l_3}C_{n_1 n_2 n_3}^{l_1 l_2 l_3}, & \label{iJps}
\eea
may be derived within the frames of the angular momentum theory 
\cite{JB77,VMK88,BL81},
when 
\[ l_a=k_a+\tfrac 12(\alpha _a+\beta _a),\quad m_a=\tfrac 12
(\alpha _a+\beta _a),\quad n_a=\tfrac 12(\beta _a-\alpha _a) \]
and 
\begin{eqnarray*}
& \alpha _a=m_a-n_a,\quad \beta _a=m_a+n_a,\quad k_a=l_a-m_a; & \\
& \alpha _3=\alpha _1+\alpha _2,\quad \beta _3=\beta _1+\beta _2 &
\end{eqnarray*}
are non-negative integers (cf.\ \cite{Vi65a}). Unfortunately, quite an 
elaborate expansion \cite{KK73,KMSS92} of two multiplied Jacobi or 
Gegenbauer polynomials in terms of the third Jacobi or Gegenbauer 
polynomial in frames of (\ref{iJps}) gives rather complicated multiple-sum
expressions for the integrals involving the ultraspherical or 
polyspherical functions in the generic SO($n$) or U($n$) case. 

In section 2, the definite integrals involving products of three 
unrestrained Jacobi polynomials are initially expressed using the direct 
(cf.\ \cite{Ga73,Ju93}) integration procedure as different (more or less 
symmetric) triple-sums in terms of beta and gamma functions. Later they 
are rearranged straightforwardly (without any allusion to special 
isofactors of Sp(4)) to a more convenient form, with smaller number of 
sums, or at least, with a richer structure of the summation intervals 
(responding to the triangular conditions of the coupling coefficients) 
and better possibilities of summation (especially, under definite 
restrictions or for some coinciding parameters). In section 3, these 
results are extended and specified for the definite integrals involving 
the triplets of the multiplied Gegenbauer polynomials and related special 
Jacobi polynomials.

In section 4, some normalization and phase choice peculiarities of 
the canonical basis states and matrix elements of the symmetric 
(class-one) irreducible representations of SO($n$) are discussed. 
Then we consider the corresponding expressions of $3j$-symbols and 
Clebsch--Gordan coefficients of SO($n$), factorized in terms of integrals 
involving triplets of the Gegenbauer polynomials (preferable in 
comparison with results of \cite{Ju93}) and extreme (summable) 
$3j$-symbols, together with the alternative phase systems.

In section 5, the semicanonical basis states and matrix elements of the
symmetric (class-one) irreducible representations of SO$(n)$ for 
restriction SO$(n)\supset $ SO$(n^{\prime })\times $SO$(n^{\prime 
\prime })$ ($n^{\prime }+n^{\prime \prime }=n$) are discussed. The 
corresponding factorized $3j$-symbols and Clebsch--Gordan coefficients, 
expressed in terms of integrals involving triplets of the Jacobi 
polynomials and extreme $3j$-symbols, are considered, together with a 
special approach to the $n^{\prime \prime }=2$ and 
$n^{\prime }=n^{\prime \prime }$ cases.

The spherical functions for the canonical chain of unitary groups 
U$(n)\!\supset $ U$(n-1)\times $U(1)$\supset \cdot \cdot \cdot \supset $%
U$(2)\times $U$(1)\supset $U$(1)$ correspond to the matrix elements of 
the class-two (mixed tensor) representations of U($n$), which include 
the scalar of subgroup U($n-1$) (see \cite{KMSS92}). The factorized 
$3j$-symbols of U($n$), related in this case to isofactors of 
SO$(2n)\!\supset $SO$(2n-2)\times $SO(2), are expressed in section 6, 
in terms of special integrals involving triplets of the Jacobi 
polynomials.

In section 7, we discuss the weight lowering (shift) operators of Sp(4) 
in its enveloping algebra and (in section 8) the analytical continuation 
relation of triple-sum series with the semistretched isoscalar factors of 
the second kind of Sp(4) \cite{AlJ71}. Particularly, some special cases 
of these isoscalar factors turn into the double or single sum series for 
some coinciding values of parameters.

In section 9, the expansions in terms of the double-sum seed isofactors 
for the most general isofactors of SO($n$) for the coupling of the two 
symmetric irreps in the canonical basis are considered.

\section{Integrals involving triplets of Jacobi polynomials}

It is convenient for our purposes to use the symmetry property and the 
two following expressions for the Jacobi polynomials (cf.\ (16) of 
section 10.8 of \cite{BE53}, chapter 22 of \cite{AS65} and definition 
2.5.1 of \cite{AAR99}): 
\begin{subequations} \bea
& {P_k^{(\alpha ,\beta )}(x)=(-1)^{k}P_k^{(\beta ,\alpha )}(-x)} & 
\label{dJps} \\
& {=2^{-k}\dsty \sum_{m}\frac{(-k-\alpha )_m(-k-\beta )_{k-m}}{m!(k-m)!}
(-1)^{m}(1+x)^{m}(1-x)^{k-m}}&  \label{dJpa} \\
& {=(-1)^{k}\dsty \sum_{m}\frac{(-k-\alpha )_m(k+\alpha +\beta +1)_{k
-m}}{m!(k-m)!}\left( \frac{1-x}{2}\right) ^{k-m},} & \label{dJpb}
\eea \end{subequations}
where $\alpha >-1,\beta >-1$ and 
\[ (c)_n=\prod_{k=0}^{n-1}(c+k)=\frac{\Gamma (c+n)}{\Gamma (c)} \]
are the Pochhammer symbols.

We introduce the following expressions for the integrals involving the 
product of three Jacobi polynomials $P_{k_1}^{(\alpha _1,\beta _1)}(x)$, 
$P_{k_2}^{(\alpha _2,\beta _2)}(x)$ and $P_{k_3}^{(\alpha _3,\beta _3)}
(x)$ with a measure dependent on $\alpha _0>-1,\beta _0>-1$ and integers 
$\alpha _a-\alpha _0\geq 0$, $\beta _a-\beta _0\geq 0$ ($a=1,2,3$): 
\begin{subequations} \bea
& {\dsty \frac 12\int\limits_{-1}^1\d x\left( \frac{1+x}{2}\right)
^{(\beta _1+\beta _2+\beta _3-\beta _0)/2}\left( \frac{1-x}{2}%
\right) ^{(\alpha _1+\alpha _2+\alpha _3-\alpha _0)/2}
\prod_{a=1}^{3}P_{k_a}^{(\alpha _a,\beta _a)}(x)} &  \nonumber \\
& {=\widetilde{\cal I}\left[ 
\begin{array}{cccc}
\alpha _0,\beta _0 & \alpha _1,\beta _1 & \alpha _2,\beta _2 & 
\alpha _3,\beta _3 \\ 
& k_1 & k_2 & k_3
\end{array} \right] } & \label{iJp}  \\
& {=(-1)^{k_1+k_2+k_3}\widetilde{\cal I}\left[ \begin{array}{cccc}
\beta _0,\alpha _0 & \beta _1,\alpha _1 & \beta _2,\alpha _2 & 
\beta _3,\alpha _3 \\ 
& k_1 & k_2 & k_3
\end{array} \right] } & \label{iJpt} \\
& {=\dsty\sum_{z_1,z_2,z_3}{\rm B}\left( 1\!-\!\tfrac 12\beta _0\!+\!%
\tsum_{a=1}^{3}(\tfrac 12\beta _a\!+\!z_a),1\!-\!\tfrac 12\alpha _0\!+\!%
\tsum_{a=1}^{3}(\tfrac 12\alpha _a\!+\!k_a\!-\!z_a)\right) } & \nonumber \\
& {\times \dsty\prod_{a=1}^{3}\frac{(-1)^{z_a}(-k_a-\alpha _a)_{z_a}(-k_a
-\beta _a)_{k_a-z_a}}{z_a!(k_a-z_a)!}} &  \label{iJpb} \\
& {=\dsty \sum_{z_1,z_2,z_3}{\rm B}\left( 1-\tfrac 12\beta _0+\tfrac 12
\tsum_{a=1}^{3}\beta _a,1-\tfrac 12\alpha _0+\tsum_{a=1}^{3}(\tfrac 12
\alpha _a+k_a-z_a)\right) } & \nonumber \\
& {\times \dsty \prod_{a=1}^{3}\frac{(-1)^{k_a}(-k_a-\alpha _a)_{z_a}
(k_a+\alpha _a+\beta _a+1)_{k_a-z_a}}{z_a!(k_a-z_a)!}} &  \label{iJpc} \\
& {=\dsty \sum_{z_1,z_2,z_3}\frac{(-1)^{k_1+k_3-z_3}(-k_1-\alpha _1)_{z_1}
(k_1+\alpha _1+\beta _1+1)_{k_1-z_1}}{z_1!(k_1-z_1)!}} & \nonumber \\
& {\times \dsty \frac{(-k_2-\beta _2)_{z_2}(k_2+\alpha _2+\beta _2
+1)_{k_2-z_2}(-k_3-\alpha _3)_{k_3-z_a}(-k_3-\beta _3)_{z_3}}{z_2!
(k_2-z_2)!z_3!(k_3-z_3)!}} & \nonumber \\
& {\times \dsty {\rm B}\left( 1\!-\!\tfrac 12\beta _0\!+\!\tfrac 12
\tsum_{a=1}^{3}\beta _a\!+\!\tsum_{b=2}^{3}(k_b\!-\!z_b),1\!-\!\tfrac 12
\alpha _0\!+\!\tfrac 12\tsum_{a=1}^{3}\alpha _a\!+\!k_1\!-\!z_1\!+\!z_3
\right)} &  \label{iJpf} \\
& {={\rm B}( k_i+\alpha _i+1,k_i+\beta _i+1)\dsty \sum_{z_1,z_2,z_3}
(-1)^{p_i-p_i^{\prime \prime }+z_1+z_2+z_3}} &  \nonumber \\
& {\times \dsty \binom{p_i^{\prime \prime }}{p_i-z_1-z_2-z_3}
\frac{(k_i+1)_{z_i}(k_i+\beta _i+1)_{z_i}}{z_i!(2k_i+\alpha _i+\beta _i
+2)_{z_i}}} & \nonumber \\
& {\times \dsty \prod_{a=j,k;a\neq i}\frac{(-k_a-\beta _a)_{z_a}(k_a
+\alpha _a+\beta _a+1)_{k_a-z_a}}{z_a!(k_a-z_a)!},} &  \label{iJpd}
\eea \end{subequations}
where the linear combinations (triangular conditions)
\begin{eqnarray*}
& {p_i^{\prime } =\tfrac 12(\beta _j+\beta _k-\beta _i-\beta _0)\geq 0,
\quad p_i^{\prime \prime }=\tfrac 12(\alpha _j+\alpha _k-\alpha _i
-\alpha _0)\geq 0,} & \\
& {p_i=k_j+k_k-k_i+p_i^{\prime }+p_i^{\prime \prime }\geq 0\quad \quad
(i,j,k=1,2,3)} &
\end{eqnarray*}
and arguments of the binomial coefficients are the non-negative integers. 
These integrals would otherwise vanish. Three first expressions 
(\ref{iJpb}), (\ref{iJpc}) and (\ref{iJpf}) (including $(k_1+1)(k_2+1)
(k_3+1)$ terms each) are derived directly using expressions (\ref{dJpa}), 
(\ref{dJpb}) and their combination, respectively, for the Jacobi 
polynomials and definite integrals (see equation (6.2.1) of \cite{AS65} 
or section 1.1 of \cite{AAR99}) in terms of beta functions 
${\rm B}(x,y)=\Gamma (x)\Gamma (y)/\Gamma (x+y)$. Two first expressions 
(\ref{iJpb}) and (\ref{iJpc}) are invariant under permutations of the 
sets $k_a,\alpha _a,\beta _a$, $a=1,2,3$, but only (\ref{iJpb}) satisfies 
relation (\ref{iJp})--(\ref{iJpt}). Expression (\ref{iJpf}) with minimal 
symmetry\footnote{It gets the phase factor $(-1)^{k_1+k_2+k_3}$ after 
interchange of the sets $k_1,\alpha _1,\beta _1,\alpha _3,\alpha _0$ and 
$k_2,\beta _2,\alpha _2, \beta _3,\beta _0$.} is derived using 
(\ref{dJpb}) for $P_{k_1}^{(\alpha _1,\beta _1)}(x)$, (\ref{dJpb}) 
together with (\ref{dJps}) for $P_{k_2}^{(\alpha _2,\beta _2)}(x)$ and 
(\ref{dJpa}) together with (\ref{dJps}) for 
$P_{k_3}^{(\alpha _3,\beta _3)}(x)$. However, the vanishing of integrals 
(\ref{iJp}) under spoiled triangular conditions is seen directly only in 
the final expression (\ref{iJpd}), which cannot be derived in a similar 
manner as (\ref{iJpb})--(\ref{iJpf}), but it has been proved in 
\cite{Al02a} after an elaborated analytical continuation procedure. 
The $i$th expression (\ref{iJpd}) is invariant under permutations of the 
sets $k_a,\alpha _a,\beta _a$, where $a=1,2,3;\,a\neq i$.

We get a simple rearrangement of (\ref{iJpc}) into (\ref{iJpd}), after 
we apply the symmetry relation (\ref{iJp})--(\ref{iJpt}) to (\ref{iJpc}) 
(i.e.\ interchange $\alpha _a$ and $\beta _a$, $a=0,1,2,3$).  
When $\alpha _0$ and $\beta _0$ are integers, the $_3F_2(1)$ type sums 
over $z_i$ in modified expressions (\ref{iJpc}) and (\ref{iJpd}) 
correspond to the CG coefficients of SU(2) with the equivalent Regge 
\cite{R58} $3\times 3$ symbols
\begin{subequations} 
\begin{equation}
\left\| \begin{array}{ccc}
k_i & k_i+\alpha _i+\beta _i & p_i-z_j-z_k \\ 
p_i^{\prime }\!+\!\beta  _i\!+\!k_j\!+\!k_k\!-\!z_j\!-\!z_k & 
p_i^{\prime \prime } & k_i+\alpha _i \\
p_i^{\prime \prime }+\alpha _i & 
p_i^{\prime }\!+\!k_j\!+\!k_k\!-\!z_j\!-\!z_k & k_i+\beta _i
\end{array} \right\|  \label{aRsc}
\end{equation}
and 
\begin{equation}
\left\| \begin{array}{ccc}
p_i-z_j-z_k & k_i & k_i+\alpha _i+\beta _i \\ 
k_i+\beta _i & p_i^{\prime \prime }+\alpha _i & 
p_i^{\prime }\!+\!k_j\!+\!k_k\!-\!z_j\!-\!z_k \\
k_i+\alpha _i & p_i^{\prime }\!+\!\beta _i\!+\!k_j\!+\!k_k\!-\!z_j\!-\!z_k 
& p_3^{\prime \prime }
\end{array} \right\| , \label{aRsd}
\end{equation}
\end{subequations}
expressed in the both cases by means of (15.1c) of Jucys and 
Bandzaitis \cite{JB77} [see also (7) of section 8.2 of \cite{VMK88}], but 
with hidden triangular conditions in the first case. For possible 
non-integer values of $\alpha _0$ and/or $\beta _0$, the doubts as to the 
equivalence of these finite $_3F_2(1)$ series may be caused by the 
absence of mutually coinciding integer parameters [$k_i$ in (\ref{iJpc}) 
and $\min (p_i-z_j-z_k,p_i^{\prime \prime })$ as triangular conditions in 
(\ref{iJpd}), respectively] restricting summation over $z_i$, unless 
equation (15.1d) of \cite{JB77} (together with possible inversion of 
summation) is used for the CG coefficient of SU(2) with Regge symbol 
(\ref{aRsd}).\footnote{The proof of relation between the corresponding 
finite $_3F_2(1)$ series in (\ref{iJpc}) and (\ref{iJpd}) based on 
composition of Thomae's transformation formulae (see 
\cite{AAR99,Sl66,GR90}) or their Whipple's specifications for single 
restricting parameter (see \cite{Sl66,RV-JR92}) is rather complicated.} 

In order to demonstrate an analogy with the proof of identity 
(2.6a)--(2.6c) of \cite{Al02a}, we may also consider rearrangement of 
(\ref{iJpf}) into (\ref{iJpd}) with the specified $i=3$. For special 
values  of parameters $k_1=0$, $p_3=k_2-k_3+p_3^{\prime }+p_3^{\prime 
\prime }$, the $_3F_2(1)$ type sums over $z_3$ in (\ref{iJpf}) and 
(\ref{iJpd}) correspond to the equivalent Regge symbols related to 
different expressions (15.1b) and (15.1c) of \cite{JB77} or (4) and (7) 
of section 8.2 of \cite{VMK88} for the CG coefficients of SU(2). Hence, 
for $k_1=z_1=0$ the double-sum expressions (\ref{iJpf}) and (\ref{iJpd}) 
are equal. This identity may be extended applying the symmetry relation 
(\ref{iJp})--(\ref{iJpt}) independently to the left-hand or to the 
right-hand side. Further, we see that the double-sum over $z_2$ and $z_3$ 
in the general triple-sum expression (\ref{iJpf}) corresponds to its 
$k_1=z_1=0$ version with parameters $\frac 12\alpha _1,p_3^{\prime \prime 
}$ and $p_3$ replaced by $\frac 12\alpha _1+k_1-z_1,p_3^{\prime \prime }
+k_1-z_1$ and $p_3-z_1$, respectively. Now it is expedient to insert the 
double-sum version of (\ref{iJpf})--(\ref{iJpd}) with interchanged 
$\alpha _a$ and $\beta _a$ in the right-hand side and corresponding 
parameters replaced by $\frac 12\alpha _1+k_1-z_1,p_3^{\prime \prime }
+k_1-z_1$ and $p_3-z_1$. Hence, we derive the final result, which is 
related to (\ref{iJpd}) with respect to the symmetry relation 
(\ref{iJp})--(\ref{iJpt}). Note that direct transformation of single 
$_3F_2(1)$ series \cite{Sl66,GR90,RV-JR92} is useless in this case. 

An advantage of our new expression (\ref{iJpd}) is the restriction of 
all three summation parameters $z_1+z_2+z_3$ by the triangular condition 
$p_i$, in contrast with remaining expressions. Alternatively, the 
linear combination of the summation parameters $p_i-z_1-z_2-z_3\geq 0$ is 
restricted in addition by $p_i^{\prime \prime }$ (or by $p_i^{\prime }$, 
if symmetry relation (\ref{iJp})--(\ref{iJpt}) is applied) only in the 
$i$th version of (\ref{iJpd}). Thus, there are three cases when the 
expressions for integral $\widetilde{\cal I}\left[ \cdot \cdot \cdot 
\right] $ are completely summable and six cases when they turn into 
double sums, in addition to the double sums which appear for $k_a=0$ 
($a=1,2,3$). However, the factorization of (\ref{iJpb})--(\ref{iJpd}) for 
$\alpha _0=\beta _0=0$, $\alpha _i=\alpha _j+\alpha _k$, 
$\beta _i=\beta _j+\beta _k$ into a product of two CG coefficients of 
SU(2) is not straightforward to prove.

For $\alpha _0=0$ and $\alpha _3=\alpha _1+\alpha _2$, the integrals 
involving the product of three Jacobi polynomials (\ref{iJpd}) turn into 
the double sums (\ref{iJra}) or (\ref{iJrb})
\begin{subequations} \bea
& {\widetilde{\cal I}\left[ \begin{array}{cccc}
0,\beta _0 & \alpha _1,\beta _1 & \alpha _2,\beta _2 & 
\alpha _1+\alpha _2,\beta _3 \\ 
& k_1 & k_2 & k_3
\end{array} \right] ={\rm B}(k_3\!+\!\alpha _1\!+\!\alpha _2\!+\!1,
k_3\!+\!\beta _3\!+\!1)} & 
\nonumber \\
& {\times \dsty \sum_{z_1,z_2}\frac{(k_3+1)_{p_3-z_1-z_2}(k_3+\beta _3
+1)_{p_3-z_1-z_2}}{(p_3-z_1-z_2)!(2k_3+\alpha _1+\alpha _2+\beta _3
+2)_{p_3-z_1-z_2}}} &  \nonumber \\
& {\times \dsty \prod_{a=1}^{2}\frac{(-k_a-\beta _a)_{z_a}(k_a+\alpha _a
+\beta _a+1)_{k_a-z_a}}{z_a!(k_a-z_a)!}} &  \label{iJra} \\
& {={\rm B}\left( k_1+\alpha _1+1,k_1+\beta _1+1\right) } &  \nonumber \\
& {\times \dsty \sum_{z_2,z_3}\binom{k_2+\alpha _2}{z_2}\frac{(-1)^{%
\alpha _2-z_2}(k_2+\alpha _2+\beta _2+1-z_2)_{k_2}(-k_3-\beta _3)_{z_3}}{%
z_3!(k_3-z_3)!k_2!}} &  \nonumber \\
& {\times \dsty \frac{(k_3+\alpha _3+\beta _3+1)_{k_3-z_3}(k_1+1)_{p_1
-z_2-z_3}(k_1+\beta _1+1)_{p_1-z_2-z_3}}{(p_1-z_2-z_3)!(2k_1+\alpha _1
+\beta _1+2)_{p_1-z_2-z_3}},} &  \label{iJrb}
\eea \end{subequations}
both related to the Kamp\'{e} de F\'{e}riet \cite{K-F21,AK-F26} functions 
$F_{2:1}^{2:2}$. It is evident that the triple series (\ref{iJpb}) and 
(\ref{iJpc}) with $\alpha _0=0$ may be also extended to the negative 
integer values of $\alpha _2$,
\bea
& {\widetilde{\cal I}\left[ \begin{array}{cccc}
0,\beta _0 & \alpha _1,\beta _1 & \alpha _2,\beta _2 & 
\alpha _3,\beta _3 \\ 
& k_1 & k_2 & k_3
\end{array} \right] } &  \nonumber \\
& {=(-1)^{\alpha _2}\dsty \frac{(k_2+\alpha _2)!(k_2+\beta _2)!}{%
k_2!(k_2+\alpha _2+\beta _2)!}\widetilde{\cal I}\left[ 
\begin{array}{cccc}
0,\beta _0 & \alpha _1,\beta _1 & -\alpha _2,\beta _2 & 
\alpha _3,\beta _3 \\ 
& k_1 & k_2+\alpha _2 & k_3
\end{array} \right] ,} &  \label{iJpe}
\eea
with invariant values of $p_1$ and $p_3$. Hence, using (\ref{iJra}) for 
the right-hand side of (\ref{iJpe}), the left-hand side of (\ref{iJpe}) 
may be expressed as the double sum for $\alpha _3=\alpha _1-\alpha _2$ 
and (\ref{iJrb}) may be derived after interchange of 
$k_1,\alpha _1,\beta _1$ and $k_3,\alpha _3,\beta _3$. 

\section{Integrals involving triplets of Gegenbauer polynomials}

The Gegenbauer (ultraspherical) polynomial $C_k^{\lambda }(\cos \theta 
)$ may be expressed as the finite series \cite{BE53,AS65}, or in terms of 
the special Jacobi polynomial (cf.\ \cite{BE53,AAR99}) 
\begin{subequations} \bea
& {C_k^{\lambda }(\cos \theta )=\dsty \sum_{m=0}^{[k/2]}\frac{(-1)^{m}
(\lambda )_{k-m}}{m!(k-2m)!}2^{k-2m}\cos ^{k-2m}\theta } &  \label{dGpa} \\
& {=\dsty \frac{(2\lambda )_k}{(\lambda +1/2)_k}P_k^{(\lambda -1/2,
\lambda -1/2)}(\cos \theta ),} & \label{dGpJ}
\eea \end{subequations}
where $[k/2]=\frac 12 (k-\delta )$ ($\delta =0$ or 1) is an integer part 
of $k/2$ and (\ref{dGpJ}) includes almost twice as many terms as 
(\ref{dGpa}).

Now we may express the integrals involving the product of three 
Gegenbauer polynomials $C_{k_i}^{\lambda _i}(x)$ as follows: 
\begin{subequations} \bea
& {\dsty \int\limits_0^{\pi }\d \theta (\sin \theta )^{\lambda _1+
\lambda _2+\lambda _3-n/2+1}\prod_{i=1}^{3}C_{k_i}^{\lambda _i}(\cos 
\theta ) } & \nonumber \\
& {=\dsty \sum_{z_1,z_2,z_3}{\rm B}\left(1+\tfrac 12\left( 
\tsum_{a=1}^{3}\lambda _a-n/2\right),\tfrac 12+\tsum_{a=1}^{3}(\tfrac 12
k_a-z_a)\right) } & \nonumber \\
& {\times \dsty \prod_{i=1}^{3}\frac{(-1)^{z_i}2^{k_i-2z_i}(\lambda _i)_{%
k_i-z_i}}{z_i!(k_i-2z_i)!}} &  \label{iGpa} \\
& {=(-1)^{[k_1/2]+[k_2/2]+[k_3/2]}\dsty \prod_{a=1}^{3}\frac{(\lambda _a%
)_{(k_a+\delta _a)/2}}{(1/2)_{(k_a+\delta _a)/2}}} &  \nonumber \\
& {\times \widetilde{\cal I}\left[ \begin{array}{cccc}
-\frac 12,\frac{n-3}{2} & \delta _1-\frac 12,\lambda _1-\frac 12 & 
\delta _2-\frac 12,\lambda _2-\frac 12 & 
\delta _3-\frac 12,\lambda _3-\frac 12 \\ 
& \frac 12(k_1-\delta _1) & \frac 12(k_2-\delta _2) & 
\frac 12(k_3-\delta _3)
\end{array} \right] } &  \label{iGpb} \\
& {=(-1)^{p_i^{\prime }/2}{\rm B}\left( \tfrac 12(k_i+\delta _i+1),
\lambda _i+\tfrac 12(k_i-\delta _i+1)\right) } &  \nonumber \\
& {\times \dsty \prod_{a=1}^{3}\frac{(\lambda _a)_{(k_a+\delta _a)/2}}{%
(1/2)_{(k_a+\delta _a)/2}}\sum_{z_1,z_2,z_3}\binom{(\delta _j+\delta _k
-\delta _i)/2}{p_i/2-z_1-z_2-z_3}} &  \nonumber \\
& {\times (-1)^{z_1+z_2+z_3}\dsty \frac{\left( (k_i-\delta _i)/2+1\right) %
_{z_i}\left( \lambda _i+(k_i-\delta _i+1)/2\right) _{z_i}}{z_i!
(k_i+\lambda _i+1)_{z_i}}} & \nonumber \\
& {\times \dsty \prod_{a\neq i}\frac{\left( -\lambda _a-(k_a-\delta _a
-1)/2\right) _{z_a}\left( \lambda _a+(k_a+\delta _a)/2\right) _{(k_a
-\delta _a)/2-z_a}}{z_a!\left( (k_a-\delta _a)/2-z_a\right) !}} &  
\label{iGpc} \\
& {=\widetilde{\cal I}\left[ \begin{array}{cccc}
\overline{\alpha }_0,\overline{\alpha }_0 & 
l_1^{\prime }+\overline{\alpha }_0,l_1^{\prime }+\overline{\alpha }_0 & 
l_2^{\prime }+\overline{\alpha }_0,l_2^{\prime }+\overline{\alpha }_0 & 
l_3^{\prime }+\overline{\alpha }_0,l_3^{\prime }+\overline{\alpha }_0  \\
& k_1 & k_2 & k_3
\end{array} \right] } &  \nonumber \\
& {\times 2^{l_1^{\prime }+l_2^{\prime }+l_3^{\prime }+n-2}\dsty
\prod_{i=1}^{3}\frac{(2\lambda _i)_{k_i}}{(\lambda _i+1/2)_{k_i}},} & 
\label{iGpd} 
\eea \end{subequations}
where in (\ref{iGpc})
\begin{eqnarray*}
& {p_i^{\prime }=\tfrac 12(\lambda _j+\lambda _k-\lambda _i-\tfrac n2
+1)\geq 0, \quad p_i^{\prime \prime }=\tfrac 12(\delta _j+\delta _k
-\delta _i),} & \\
& {p_i=\tfrac 12(k_j+k_k-k_i)+p_i^{\prime }\geq 0 } &
\end{eqnarray*}
and in (\ref{iGpd}) $l_a^{\prime }=\lambda _a-\tfrac n2+1$, 
$\overline{\alpha }_{0}=\tfrac{n-3}{2}$. In accordance with (\ref{iGpc}), 
these integrals would otherwise vanish. Expressions (\ref{iGpa}) (cf.\ 
\cite{Ju93}) and (\ref{iGpd}) (cf.\ \cite{Ga73}) are derived directly 
(using definite integrals (6.2.1) of \cite{AS65} or \cite{AAR99} in terms 
of beta functions). Further (\ref{iGpa}) is recognized as consistent with 
a particular case of (\ref{iJpc})\footnote{This was the reason why 
(\ref{iJpc}) has been introduced in \cite{Al02a}.} denoted by 
(\ref{iGpb}) and re-expressed, in accordance with 
(\ref{iJpc})--(\ref{iJpd}), in the most convenient form as (\ref{iGpc}), 
where $\delta _1,\delta _2,\delta _3=0$ or 1 (in fact either 
$\delta _1=\delta _2=\delta _3=0$, or $\delta _a=\delta _b=1$, 
$\delta _c=0$) and $\frac 12(k_a-\delta _a)$ ($a=1,2,3$) are integers.

Expression (\ref{iGpa}) includes $\frac 18\prod_{a=1}^{3}(k_a-\delta _a
+2)$ terms, when (\ref{iGpd}), used together with (\ref{iJpb}) or 
(\ref{iJpc}), each includes $\prod_{a=1}^{3}(k_a+1)$ terms; otherwise, 
the number of terms in the $i$th version of the most convenient formula 
(\ref{iGpc}) never exceeds 
\begin{equation}
A_i=(p_i^{\prime \prime }+1)\min \left[ \tfrac 12(p_i+1)(p_i-
p_i^{\prime \prime }+2),\tfrac 14\tprod_{a\neq i}(k_a-\delta _a+2)\right] ,
\label{ntc}
\end{equation}
where a set $i,j,k$ is a transposition of 1,2,3. The number of terms 
decreases in comparison with (\ref{ntc}) in the intermediate region 
\[ \tfrac 12\min (k_j-\delta _j,k_k-\delta _k)<p_i<\tfrac 12(k_j-
\delta _j+k_k-\delta _k). \]

Actually, expression (\ref{iGpc}) is related to the Kamp\'{e} de 
F\'{e}riet \cite{K-F21,AK-F26} function $F_{2:1}^{2:2}$ (for 
$p_i^{\prime \prime }=0$) or to the sum of two such functions (when 
$p_i^{\prime \prime }=1$). Hence, after comparing three different 
versions of (\ref{iGpc}), the rearrangement formulae of special 
Kamp\'{e} de F\'{e}riet functions $F_{2:1}^{2:2}$ can be derived.

Finally we may express the integrals involving the product of three 
Gegenbauer polynomials $C_{l_1-l_1^{\prime }}^{l_1^{\prime }+n/2-1}(x)$, 
$C_{l_2-l_2^{\prime }}^{l_2^{\prime }+n/2-1}(x)$ and 
$C_{l_3-l_3^{\prime }}^{l_3^{\prime }+n/2-1}(x)$, needed in the next 
section as follows: 
\bea
& {\dsty \int\limits_0^{\pi }\d \theta (\sin \theta )^{l_1^{\prime
}+l_2^{\prime }+l_3^{\prime }+n-2}\prod_{i=1}^{3}C_{l_i-l_i^{%
\prime }}^{l_i^{\prime }+n/2-1}(\cos \theta ) } & \nonumber \\
& {=(-1)^{(l_j^{\prime }+l_k^{\prime }-l_i^{\prime })/2}{\rm B}%
\left( \tfrac 12(l_i-l_i^{\prime }+\delta _i+1),\tfrac 12%
(l_i+l_i^{\prime }-\delta _i+n-1)\right) } &  \nonumber \\
& {\times \dsty \prod_{a=1}^{3}\frac{(l_a^{\prime }+n/2-1)_{(l_a-
l_a^{\prime }+\delta _a)/2}}{(1/2)_{(l_a\!-\!l_a^{\prime }\!+\!\delta _a%
)/2}}\sum_{z_1,z_2,z_3}\binom{(\delta _j+\delta _k-\delta _i)/2}{(l_j\!%
+\!l_k\!-\!l_i)/2\!-\!z_1\!-\!z_2\!-\!z_3}} &  \nonumber \\
& {\times \dsty \prod_{a\neq i}\frac{\left( -(l_a\!+\!l_a^{\prime %
}\!-\!\delta _a\!+\!n\!-\!3)/2\right) _{z_a}\left( (l_a\!+\!l_a^{\prime %
}\!+\!\delta _a\!+\!n)/2\!-\!1\right) _{(l_a-l_a^{\prime }-\delta _a)/2
-z_a}}{z_a!\left( (l_a-l_a^{\prime }-\delta _a)/2-z_a\right) !}} &  
\nonumber \\
& {\times (-1)^{z_1+z_2+z_3}\dsty \frac{\left( (l_i-l_i^{\prime }
-\delta _i)/2+1\right) _{z_i}\left( (l_i+l_i^{\prime }-\delta _i+
n-1)/2\right) _{z_i}}{z_i!(l_i+n/2)_{z_i}},} &  \label{iGpco} 
\eea 
where $p_i=\frac 12(l_j+l_k-l_i)\geq 0$ and $p_i^{\prime }=\frac 12
(l_j^{\prime }+l_k^{\prime }-l_i^{\prime })\geq 0$ are integers.

Now we consider more specified integrals involving several Gegenbauer
polynomials. At first, using (\ref{iGpco}) with $i=3$ and $z_2=\delta
_2=0,\;z_3=\frac 12(l_1+l^{\prime }-l_3)-z_1$, we take special integral 
involving two multiplied Gegenbauer polynomials (where third trivial 
polynomial $C_0^{l^{\prime }+n/2-1}(x)=1$ may be inserted) in terms of 
the summable balanced (Saalsch\"{u}tzian) $_3F_2(1)$ series (cf.\ 
\cite{Sl66,GR90}) and write: 
\begin{subequations} \bea
& {\dsty \int\limits_0^{\pi }(\sin \theta )^{2l^{\prime }+n-2}C_{l_1
-l^{\prime }}^{l^{\prime }+n/2-1}(\cos \theta )C_{l^{\prime }-l^{\prime 
}}^{l^{\prime }+n/2-1}(\cos \theta )C_{l_3}^{n/2-1}(\cos \theta )
\d \theta } & \nonumber \\
& {=\dsty \int\limits_0^{\pi }(\sin \theta )^{2l^{\prime }+n-2}C_{l_1
-l^{\prime }}^{l^{\prime }+n/2-1}(\cos \theta )C_{l_3}^{n/2-1}(\cos 
\theta )\d \theta } & \label{iGp2d} \\
& {=\dsty \frac{(-1)^{(l_3-l_1+l^{\prime })/2}\pi \,l^{\prime })!(l_1
+l^{\prime }+n-3)!}{2^{2l^{\prime }+n-3}(l_1-l^{\prime })!
(J^{\prime }-l_1)!(J^{\prime }-l_3)!\;\Gamma (n/2-1)}} &  \nonumber \\
& {\times \dsty \frac{\Gamma (J^{\prime }-l^{\prime }+n/2-1)}{\Gamma 
(l^{\prime }+n/2-1)\Gamma (J^{\prime }+n/2)},} &  \label{iGp2}
\eea \end{subequations}
where $J^{\prime }=\tfrac 12(l_1+l^{\prime }+l_3)$.

Using the expansion formula (linearization Theorem 6.8.2 of \cite{AAR99}) 
of two multiplied Gegenbauer polynomials (zonal spherical functions) 
$C_{l}^{p}(x)C_k^{p}(x)$ in terms of the third polynomial $C_n^{p}(x)$ 
(cf.\ \cite{Vi65a}), where $l+k-n$ is even, the special integral 
involving three Gegenbauer polynomials (with coinciding superscripts in 
two cases) may be also expanded in terms of integrals (\ref{iGp2}) and 
may be presented as follows:
\begin{subequations} \bea
& {\dsty \int\limits_0^{\pi }(\sin \theta )^{2l^{\prime }+n-2}C_{l_1
-l^{\prime }}^{l^{\prime }+n/2-1}(\cos \theta )C_{l_2-l^{\prime }}^{%
l^{\prime }+n/2-1}(\cos \theta )C_{l_3}^{n/2-1}(\cos \theta )\d 
\theta } & \nonumber \\
& {=\dsty \frac{\pi l^{\prime }!}{2^{2l^{\prime }+n-3}\Gamma ^{3}(n/2-1)
\Gamma (l^{\prime }+n/2-1)}} &  \nonumber \\
& {\times \dsty \sum_{k=|l_1-l_2|+l^{\prime }}^{l_1+l_2-l^{\prime }}
\frac{(-1)^{(l_3+l^{\prime }-k)/2}(k+n/2-1)}{\nabla ^{2}\left( %
l^{\prime }/2,l_3/2+n/4-1,k/2+n/4-1\right) }} &  \nonumber \\
& {\times \dsty \frac{\nabla ^{2}\left( (l_1+l^{\prime }+n)/2-2,l_2/2
+n/4-1,k/2+n/4-1\right) }{\nabla ^{2}\left( (l_1-l^{\prime })/2,l_2/2
+n/4-1,k/2+n/4-1\right) }} &  \label{iGpk} \\
& {=\dsty \frac{\pi \,l^{\prime }!\prod_{a=1}^{3}\Gamma (J-l_a+n/2-1)}{%
2^{2l^{\prime }+n-3}\Gamma (n/2-1)\Gamma (l^{\prime }+n/2-1)\Gamma 
(J+n/2)}} &  \nonumber \\
& {\times \dsty \sum_{u}\frac{(-1)^{u}(J+l^{\prime }+n-3-u)!}{u!(l^{%
\prime }-u)!(J-l_1-u)!(J-l_2-u)!(J-l_3-l^{\prime }+u)!}} &  \nonumber \\
& {\times [\Gamma (n/2-1+u)\Gamma (l^{\prime }+n/2-1-u)]^{-1},} & 
\label{iGpr}
\eea \end{subequations}
where $J=\tfrac 12(l_1+l_2+l_3)$ and the gamma functions under
summation sign in the intermediate formula (\ref{iGpk}) (which is 
equivalent to (15) of \cite{HJu99}) are included into the asymmetric 
triangle coefficients 
\begin{subequations} \bea
\nabla (abc)=\left[ \frac{(a+b-c)!(a-b+c)!(a+b+c+1)!}{(b+c-a)!}
\right] ^{1/2}  \label{nabla} \\
=\left[ \frac{\Gamma (a+b-c+1)\Gamma (a-b+c+1)\Gamma (a+b+c+2)}{%
\Gamma (b+c-a+1)}\right] ^{1/2}.  \label{nablg}
\eea \end{subequations}
Finally, the sum in (\ref{iGpk}) corresponds to a very well-poised 
$_7F_6(1)$ hypergeometric series (which may be rearranged using 
Whipple's transformation of \cite{AAR99} or (6.10) of \cite{LB94} into 
balanced terminating $_4F_3(1)$ hypergeometric series) or to the usual 
$6j$ coefficient of SU(2)
\begin{equation}
\left\{ 
\begin{array}{ccc}
l^{\prime }+\frac 12n-2 & \frac 12(l_1+n)-2 & \frac 12l_1 \\ 
\frac 12l_3+\frac{1}{4}n-1 & \frac 12l_2+\frac{1}{4}n-1 & 
\frac 12l_2+\frac{1}{4}n-1
\end{array}
\right\}   \label{csR}
\end{equation}
with standard (integer or half-integer) parameters (for $n$ even), in 
accordance with expression (C3) of the $6j$ coefficient \cite{Al92} in 
terms of (\ref{nabla}). For $n$ odd some of its parameters may be 
quartervalued (i.e.\ multiple of 1/4). Using the most symmetric (Racah) 
expression \cite{JB77,VMK88} for (\ref{csR}), the final expression 
(\ref{iGpr}) with single sum is derived. Intervals of summation are 
restricted by $\min (l^{\prime },J-l_1,J-l_2,J-l_3)$ and, of course, 
(\ref{iGpr}) coincide with result of Vilenkin \cite{Vi65a} for 
$l^{\prime }=0$. Nevertheless, the use of less symmetric expressions 
(29.1b) and (29.1c) of Jucys and Bandzaitis \cite{JB77} (see also (5) 
and (6) in section 9.2 of \cite{VMK88}) for $6j$ coefficients (\ref{csR}),
together with Dougall's summation formula \cite{Sl66} of the very 
well-poised series, allowed us \cite{Al02b} to derive expressions of 
6$j$-symbols for symmetric representations of SO($n$) as the double 
series.

Comparing expansion (\ref{iGpc}) of the integrals involving triplets of 
the Gegenbauer polynomials with (\ref{iGpd}), we may write an expression 
for the integrals involving triplets of special Jacobi polynomials, with 
mutually equal superscripts, 
\bea
& \widetilde{\cal I}\left[ \begin{array}{cccc}
\alpha _0,\alpha _0 & \alpha _1,\alpha _1 & 
\alpha _2,\alpha _2 & \alpha _3,\alpha _3 \\ 
& k_1 & k_2 & k_3
\end{array} \right] & \nonumber \\
& {=\dsty \frac{[1+(-1)^{p_i}]\,{\rm B}(1/2,k_i+\alpha _i+1)}{2^{k_1+
k_2+k_3+\alpha _1+\alpha _2+\alpha _3-\alpha _0+2}(1/2)_{(k_j+\delta _j)
/2}(1/2)_{(k_{k}+\delta _k)/2}}} & \nonumber \\
& {\times \dsty \sum_{z_1,z_2,z_3}(-1)^{p_i^{\prime }+(k_j+\delta
_j+k_{k}+\delta _k)/2+z_1+z_2+z_3}\binom{(\delta _j+\delta
_k-\delta _i)/2}{p_i/2-z_1-z_2-z_3}} &  \nonumber \\
& {\times \dsty \prod_{a=j,k;a\neq i}\frac{\left( -k_a\!-\!\alpha _a
\right) _{(k_a+\delta _a)/2+z_a}\left( \alpha _a+(k_a+\delta
_a\!+\!1)/2\right) _{(k_a\!-\!\delta _a)/2\!-\!z_a}}{z_a!\left( (k_a
-\delta _a)/2-z_a\right) !}} &  \nonumber \\
& {\times \dsty \binom{(k_i-\delta _i)/2+z_i}{z_i}\frac{\left( \alpha _i
+(k_i-\delta _i)/2+1\right) _{z_i}}{(\alpha _i+k_1+3/2)_{z_i}}.} &  
\label{iJpst}
\eea
Here 
\begin{eqnarray*}
& {p_i=k_j+k_{k}-k_i+p_i^{\prime }+p_i^{\prime \prime },\quad 
\tfrac 12(k_i-\delta _i),\quad \delta _i=0\;{\rm or}\;1,} & \\
& {p_i^{\prime }=p_i^{\prime \prime }=\tfrac 12(\alpha _j+\alpha
_k-\alpha _i-\alpha _0)\quad \quad(i,j,k=1,2,3)} &
\end{eqnarray*}
are non-negative integers.

Comparing expansion (\ref{iGpr}) of the integrals involving more 
specified triplets of the Gegenbauer polynomials with (\ref{iGpd}), we 
may also write an expression for the integrals involving triplets of 
special Jacobi polynomials, 
\bea
& \widetilde{\cal I}\left[ \begin{array}{cccc}
\alpha _0,\alpha _0 & \alpha _1,\alpha _1 & 
\alpha _1,\alpha _1 & \alpha _0,\alpha _0 \\ 
& k_1 & k_2 & k_3
\end{array} \right] & \nonumber \\
& {=\dsty \frac{[1+(-1)^{p_1}]2^{2\alpha _0-2}(\alpha _1-\alpha _0)!
\Gamma(\alpha _1+1/2)\prod_{a=1}^{3}\Gamma (p_a/2+\alpha _0
+1/2)}{\Gamma (1/2)\Gamma \left( (k_1+k_2+k_3)/2+\alpha _1
+3/2\right) }} &  \nonumber \\
& {\times \dsty \frac{\Gamma (\alpha _1+1+k_1)\Gamma (\alpha _1+1+k_2)
\Gamma (\alpha _0+1+k_3)}{\Gamma (2\alpha _1+1+k_1)\Gamma (2
\alpha _1+1+k_2)\Gamma (2\alpha _0+1+k_3)}} &  \nonumber \\
& {\times \dsty \sum_{u}\frac{(-1)^{u}\left( (k_1+k_2+k_3)/2+2\alpha _1
-u\right) !}{u!(\alpha _1-\alpha _0-u)!(p_1/2-u)!(p_2/2-u)!
(p_3/2+\alpha _0-\alpha _1+u)!}} &  \nonumber \\
& {\times [\Gamma (\alpha _0+1/2+u)\Gamma (\alpha _1+1/2-u)]^{-1}} & 
\label{iJpsr}
\eea
in terms of the balanced (Saalsch\"{u}tzian) $_4F_3(1)$ type series 
\cite{Sl66,GR90}. Here 
\[ p_i=k_j+k_k-k_i+2p_i^{\prime },\quad p_1^{\prime }=
p_2^{\prime }=0,\quad p_3^{\prime }=\alpha _1-\alpha _0 \]
are integers.

\section{Canonical basis states and coupling coefficients of SO($n$)}

The canonical basis states of the symmetric (class-one) irreducible
representation $l=l_{(n)}$ for the chain SO$(n)\!\supset $SO$(n-1)\supset
\cdot \cdot \cdot \supset $SO(3)$\supset $SO(2) are labelled by the 
$(n-2)$-tuple $M=(l_{(n-1)},N)=(l_{(n-1)},...,l_{(3)},m_{(2)})$ of 
integers 
\begin{equation}
l_{(n)}\geq l_{(n-1)}\geq ...\geq l_{(3)}\geq |m_{(2)}|.  \label{cnl}
\end{equation}
The dimension of representation space is 
\begin{equation}
d_l^{(n)}=\frac{(2l+n-2)(l+n-3)!}{l!(n-2)!}.  \label{dimo}
\end{equation}

Special matrix elements $D_{M0}^{n,l}(g)$ of SO($n$) irreducible
representation $l_{(n)}=l$ with zero for the $(n-2)$-tuple (0,...,0) 
depend only on the rotation (Euler) angles $\theta _{n-1},\theta _{n-2},%
...,\theta _2,\theta _1$ (coordinates on the unit sphere $S_{n-1}$) 
and may be factorized as 
\begin{equation}
D_{M0}^{n,l}(g)=t_{l^{\prime }0}^{n,l}(\theta _{n-1})D_{N0}^{n-1,
l^{\prime }}(g^{\prime }).  \label{dmel}
\end{equation}
Here $D_{N0}^{n-1,l^{\prime }}(g^{\prime })$ are the matrix elements of 
SO($n-1$) irrep $l_{(n-1)}=l^{\prime }$ (with coordinates on the unit 
sphere $S_{n-2}$). Special matrix elements of SO($n$) ($n>3$) irreducible
representation $l_{(n)}=l$ with the SO($n-1$) irrep labels 
$l_{(n-1)}=l^{\prime }$ and 0 and SO($n-2$) label $l_{(n-2)}=0$ for 
rotation with angle $\theta _{n-1}$ in the $(x_n,x_{n-1})$ plane are 
written in terms of the Gegenbauer polynomials as follows: 
\bea
& {t_{l^{\prime }0}^{n,l}(\theta _{n-1})=\dsty \left[ \frac{l!(l-l^{\prime
})!(n-3)!(l^{\prime }+n-4)!(2l^{\prime }+n-3)}{l^{\prime }!(l+l^{\prime
}+n-3)!(l+n-3)!}\right] ^{1/2}} &  \nonumber \\
& {\times (n/2-1)_{l^{\prime }}2^{l^{\prime }}\sin ^{l^{\prime }}\theta
_{n-1}C_{l-l^{\prime }}^{l^{\prime }+n/2-1}(\cos \theta _{n-1}),} &
\label{tmelc}
\eea
(see \cite{Vi65a}). Function (\ref{tmelc}) corresponds to the wavefunction
$\Psi _{k,l^{\prime }}^{c}(\theta )\!=\!\Psi _{l-l^{\prime },l^{\prime 
}}^{l+(n-3)/2}(\theta )$ of the tree technique (of the type 2b, see (2.4) 
of \cite{IPSW01}) with factor 
\[ \left[ \frac{\Gamma ((n-1)/2)\sqrt{\pi }\,d_{l^{\prime }}^{(n-1)}}{%
\Gamma (n/2)\,d_l^{(n)}}\right] ^{1/2}, \]
for appropriate normalization in the case of integration over the group
volume ($0\leq \theta \leq \pi $) with measure ${\rm B}^{-1}\left(
(n-1)/2,1/2\right) \sin ^{n-2}\theta \d \theta $. The remaining 
Euler angles are equal to 0 for the matrix element (\ref{tmelc}). In the 
case of SO(3), we obtain 
\begin{equation}
D_{m0}^{3,l}(\theta _2,\theta _1)=(-1)^{(l^{\prime }-m)/2}
t_{l^{\prime }0}^{3,l}(\theta _2)\e ^{\i m\theta _1},\;l^{\prime }=|m|,  
\label{tmelc3}
\end{equation}
in accordance with the relation \cite{Vi65a} between the associated 
Legendre polynomials $P_l^{m}(x)$ and special Gegenbauer polynomials 
$C_{l-m}^{m+1/2}(x)$ and the behavior of $P_l^{m}(x)$ under the 
reflection of $m$.

The corresponding $3j$-symbols for the chain SO$(n)\!\supset $SO$(n-1)
\supset \cdot \cdot \cdot \supset $SO(3) $\supset $SO(2) (denoted by 
brackets with simple subscript $n$ and labelled by sets 
$M_a=(l_a^{\prime },N_a)$) may be factorized as follows: 
\begin{subequations} \bea 
& {\left( \begin{array}{ccc}
l_1 & l_2 & l_3 \\ 
M_1 & M_2 & M_3
\end{array} \right) _{\!n}} & \nonumber \\
& {=\left( \begin{array}{ccc}
l_1 & l_2 & l_3 \\ 0 & 0 & 0
\end{array}
\right) _{\!n}^{-1}\dsty \int\limits_{{\rm SO}(n)}\d %
gD_{M_10}^{n,l_1}(g)D_{M_20}^{n,l_2}(g)D_{M_30}^{n,l_3}(g)} &
\label{c3jci} \\
& {=\left( \begin{array}{ccc}
l_1 & l_2 & l_3 \\ 
l_1^{\prime } & l_2^{\prime } & l_3^{\prime }
\end{array} \right) _{\!(n:n-1)}\left( \begin{array}{ccc}
l_1^{\prime } & l_2^{\prime } & l_3^{\prime } \\ 
N_1 & N_2 & N_3
\end{array} \right) _{\!n-1}.} &  \label{c3jcf}
\eea \end{subequations}
Here the isoscalar factors of $3j$-symbol for the restriction SO$(n)
\supset $SO($n-1)$ are denoted by brackets with composite subscript 
$(n:n-1)$ and are expressed in terms of integrals (\ref{iGpco}) 
involving the triplets of the Gegenbauer polynomials, 
\bea
& {\left( \begin{array}{ccc}
l_1 & l_2 & l_3 \\ 
l_1^{\prime } & l_2^{\prime } & l_3^{\prime }
\end{array} \right) _{\!(n:n-1)}=\left( \begin{array}{ccc}
l_1 & l_2 & l_3 \\ 0 & 0 & 0
\end{array} \right) _{\!n}^{-1}\left( \begin{array}{ccc}
l_1^{\prime } & l_2^{\prime } & l_3^{\prime } \\ 
0 & 0 & 0
\end{array} \right) _{\!n-1}} &  \nonumber \\
& {\times \dsty \left[ \frac{\Gamma \left( (n-1)/2\right) }{\pi ^{5/2}
\Gamma (n/2)}\right] ^{1/2}\prod_{a=1}^{3}{\cal N}_{l_a;l_a^{\prime },
\delta _a}^{(n:n-1)}\!\left[ \frac{d_{l_a^{\prime }}^{(n-1)}}{d_{l_a}^{%
(n)}}\right] ^{1/2}} &  \nonumber \\
& {\times \dsty \int\limits_0^{\pi }\d \theta (\sin \theta )^{l_1^{\prime }
+l_2^{\prime }+l_3^{\prime }+n-2}\prod_{i=1}^{3}C_{l_i-l_i^{\prime 
}}^{l_i^{\prime }+n/2-1}(\cos \theta ),} &  \label{isfc}
\eea
where 
\begin{equation}
{\cal N}_{l_a;l_a^{\prime },\delta _a}^{(n:n-1)}=
2^{l_a^{\prime }+n/2-2}\Gamma (l_a^{\prime }+n/2-1) 
\left[ \frac{(l_a-l_a^{\prime })!(2l_a+n-2)}{(l_a+l_a^{\prime }
+n-3)!}\right] ^{1/2}  \label{ncc}
\end{equation}
are normalization factors and particular $3j$-symbols 
\begin{subequations} \bea
& {\left( \begin{array}{ccc}
l_1 & l_2 & l_3 \\ 0 & 0 & 0
\end{array} \right) _{\!n} =(-1)^{\psi _n}\dsty \frac{1}{\Gamma (n/2)}  
\left[ \frac{(J+n-3)!}{(n-3)!\;\Gamma (J+n/2)}\right. } & \nonumber \\
& {\times \dsty \left. \prod_{i=1}^{3}\frac{\left( l_i+n/2-1\right) 
\Gamma (J-l_i+n/2-1)}{d_{l_i}^{(n)}(J-l_i)!}\right] ^{1/2}} & 
\label{isf0} \\
& {=(-1)^{\psi _n}\dsty \frac{\widetilde{\nabla }_{n[0,1,2,3]}^{-1}
(l_1,l_2;l_3,0)}{\Gamma (n/2)[(n-3)!]^{1/2}}\left[ \prod_{i=1}^{3}
\frac{l_i+n/2-1}{d_{l_i}^{(n)}}\right] ^{1/2}} & \label{isf0n}
\eea \end{subequations}
(vanishing for $J=\frac 12(l_1+l_2+l_3)$ half-integer) are derived in 
\cite{Ju93} (see also special Clebsch--Gordan coefficients 
\cite{KK73,NAl74a,Al87}). The triangular coefficients 
$\widetilde{\nabla }_{n[0,1,2,3]}(\cdot \cdot \cdot )$ in 
(\ref{isf0n}) are expressed as follows:
\bea
& {\widetilde{\nabla }_{n[0,1,2,3]}(a,b;e,f)} & \nonumber \\
& {=\dsty \left[ \frac{\left( \frac 12(a-b+e-f)\right) !\left( \frac 12
(b-a+e-f)\right) !}{\Gamma \left( \frac 12(a-b+e+f+n)-1\right) 
\Gamma \left( \frac 12(b-a+e+f+n)-1\right) } \right. } & \nonumber \\
& {\times \dsty \left. \frac{\left( \frac 12(a+b-e-f)\right) !\;\Gamma 
\left( \frac 12(a+b+e-f+n)\right) }{\Gamma \left( \frac 12(a+b-e+f+n)-1
\right) \left( \frac 12(a+b+e+f)+n-3\right) !}\right] ^{1/2}} & 
\label{nabls}
\eea
and in general
\begin{equation}
\widetilde{\nabla }_{n[i_1...i_k]}(a,b;e,f)=\left( \tprod_{i=0}^{7}%
\widetilde{A}_i\right) ^{1/2}\left( \tprod_{i\to [i_1...i_k]}
\widetilde{A}_i\right) ^{-1}, \label{nabl0}
\end{equation}
\begin{equation}
\begin{array}{ll}
\widetilde{A}_0=\left( \frac 12(a+b+e+f)+n-3\right) !, & 
\widetilde{A}_4=\Gamma \left( \frac 12(a+b+e-f+n)\right)\!, \\
\widetilde{A}_1=\Gamma \left( \frac 12(b-a+e+f+n)-1\right)\!,\;\;  & 
\widetilde{A}_5=\left( \frac 12(b-a+e-f)\right) !, \\
\widetilde{A}_2=\Gamma \left( \frac 12(a-b+e+f+n)-1\right)\!, & 
\widetilde{A}_6=\left( \frac 12(a-b+e-f)\right) !, \\
\widetilde{A}_3=\Gamma \left( \frac 12(a+b-e+f+n)-1\right)\!, & 
\widetilde{A}_7=\left( \frac 12(a+b-e-f)\right) !.
\end{array} \label{argn}
\end{equation}

Equation (\ref{isfc}) together with (\ref{iGpa}) is equivalent to the 
result of \cite{Ju93}, but its most convenient form is obtained%
\footnote{Note, that the factors under the square root form the rational 
numbers, taking into account that $\Gamma (1/2)=\sqrt{\pi }$.} when 
the special integral is expressed by means of double-sum expression 
(\ref{iGpco}) (for $i=1,2$ or 3, minimizing (\ref{ntc})), which ensures 
its finite rational structure for the fixed shift $\frac 12(l_1+l_2-l_3)$ 
of parameters. In the case of $l_i^{\prime }=0$, expression (\ref{iGpr}) 
for the special integral is more convenient in accordance with 
\cite{Al87}. In (\ref{isf0}), $J-l_i$ ($i=1,2,3$) and $J$ are non-negative
integers and $\psi _3=J$, in accordance with the angular momentum theory 
\cite{JB77,VMK88}. We may take also 
\begin{equation}
\psi _n=J  \label{psn}
\end{equation}
(see \cite{Ju93}) for $n\geq 4$, in order to obtain the isofactors 
(\ref{isfc}) positive where the maximal values of parameters 
$l_1^{\prime }=l_1,l_2^{\prime }=l_2,l_3^{\prime }=l_3$.

Only by taking into account the phase factor $(-1)^{(l^{\prime }-m)/2}$ of 
(\ref{tmelc3}), we can obtain the consistent signs of the usual Wigner 
coefficients ($3j$-symbols) 
\begin{equation}
\left( \begin{array}{ccc}
l_1 & l_2 & l_3 \\ 
l_1^{\prime } & l_2^{\prime } & l_3^{\prime }
\end{array} \right) _{\!(3:2)}=\left( \begin{array}{ccc}
l_1 & l_2 & l_3 \\ 
m_1 & m_2 & m_3
\end{array} \right) ,  \label{w3jc}
\end{equation}
of SO(3) or SU(2) (where $l_a^{\prime }=|m_a|$ and $m_1+m_2+m_3=0$), with 
\begin{equation}
\left( \begin{array}{ccc}
l_1^{\prime } & l_2^{\prime } & l_3^{\prime } \\ 
0 & 0 & 0
\end{array}
\right) _{\!2}=\delta _{\max (l_1^{\prime },l_2^{\prime },l_3^{\prime 
}),(l_1^{\prime }+l_2^{\prime }+l_3^{\prime })/2}(-1)^{%
(l_1^{\prime }+l_2^{\prime }+l_3^{\prime })/2}  \label{isf0o}
\end{equation}
consequently appearing in (\ref{isfc}) for $n=3$.\footnote{Of course, 
in this case the usual expressions \cite{Vi65a,JB77,VMK88,BL81} of the 
Clebsch--Gordan or Wigner coefficients of SU(2) are more preferable in 
comparison with equation (\ref{isfc}).}

We may write (cf.\ \cite{Ju93}) the following dependence between special
Clebsch--Gordan coefficients (denoted by square brackets with subscript) 
and $3j$-symbols of SO($n$): 
\begin{subequations} \bea
& {\left[ \begin{array}{ccc}
l_1 & l_2 & l_3 \\ 
M_1 & M_2 & M_3
\end{array} \right] _{\!n}\left[ \begin{array}{ccc}
l_1 & l_2 & l_3 \\ 
0 & 0 & 0
\end{array} \right] _{\!n}} &  \nonumber \\
& {=d_l^{(n)}\dsty \int\limits_{{\rm SO}(n)}\d gD_{M_10}^{n,
l_1}(g)D_{M_20}^{n,l_2}(g)\overline{D_{M_30}^{n,l_3}(g)}} &  
\label{cgci} \\
& {=d_l^{(n)}(-1)^{l_3-m_3}\left( \begin{array}{ccc}
l_1 & l_2 & l_3 \\ 
M_1 & M_2 & \overline{M_3}
\end{array} \right) _{\!n}\left( \begin{array}{ccc}
l_1 & l_2 & l_3 \\ 0 & 0 & 0
\end{array} \right) _{\!n},} &  \label{cgcw}
\eea \end{subequations}
where the $(n-2)$-tuple $\overline{M_3}$ is obtained from the 
$(n-2)$-tuple $M_3$ after reflection of the last parameter $m_3$. 
Then in the phase system with $\psi _n=J$, we obtain the following 
relation for the isofactors of CG coefficients in the canonical basis:
\begin{equation}
\left[ \begin{array}{ccc}
l_1 & l_2 & l_3 \\ 
l_1^{\prime } & l_2^{\prime } & l_3^{\prime }
\end{array} \right] _{\!(n:n-1)}=(-1)^{l_3-l_3^{\prime }}\left[ 
\frac{d_{l_3}^{(n)}}{d_{l_3^{\prime }}^{(n-1)}}\right] ^{1/2}\left( 
\begin{array}{ccc}
l_1 & l_2 & l_3 \\ 
l_1^{\prime } & l_2^{\prime } & l_3^{\prime }
\end{array} \right) _{\!(n:n-1)},  \label{isfcg}
\end{equation}
which, together with (\ref{isfc}), (\ref{isf0}), (\ref{psn}) and 
(\ref{iGpco}) or (\ref{iGpb}) substituted by (\ref{iJpb}), allows us 
to obtain the expressions for isofactors of SO$(n)\!\supset $SO($n-1$) 
derived in \cite{NAl74a} and satisfying the same phase conditions.

However, as it has been noted in \cite{NAl74b}, the choice (\ref{psn}) of 
$\psi _n$ does not give the correct phases for special isofactors of 
SO(4) \cite{B61} in terms of $9j$ coefficients of SU(2) \cite{JB77} 
and for isofactors of SO(5)$\supset $SO(4), as considered in 
\cite{AlJ71,AlJ69,H65}. The contrast of the phases is caused by the 
fact that the signs of the matrix elements of infinitesimal operators 
\[ A_{k,k-1}=x_k\frac{\partial }{\partial x_{k-1}}-x_{k-1}
\frac{\partial }{\partial x_k},\quad k=3,...,n, \]
(with exception of $A_{2,1}$) between the basis states \cite{Vi65a} of 
SO($n$) in terms of Gegenbauer polynomials (in $x_k/r_k$, 
$r_k^2=x_1^2+...+x_k^2$ variables) are opposite to the signs of the 
standard (Gel'fand--Tsetlin) matrix elements \cite{GC50b,BR77,Kl79}.
We eliminate this difference of phases and our results match with the 
isofactors for decomposition of the general and vector irreps 
$m_n \otimes 1$ of SO($n$) \cite{Kl79,GKl77} (specified also in 
\cite{PC88,SR98}) after we multiply isofactors of CG coefficients for the 
restriction SO$(n)\!\supset $SO($n-1$) ($n\geq 4$), i.e.\ the left-hand 
side of (\ref{isfcg}), by 
\[ (-1)^{(l_1+l_2-l_3-l_1^{\prime }-l_2^{\prime }
+l_3^{\prime })/2} \]
(cf.\ \cite{NAl74b}), i.e.\ after we omit the phase factors $(-1)^{\psi
_{n}}$ and $(-1)^{\psi _{n-1}}$ in the both auxiliary $3j$-symbols of 
(\ref{isfc}), as well as $(-1)^{l_3-l_3^{\prime }}$ in relation 
(\ref{isfcg}), again keeping the isofactors (\ref{isfcg}) with the 
maximal values of parameters $l_1^{\prime }=l_1,l_2^{\prime }=
l_2,l_3^{\prime }=l_3$ for this restriction positive. In the both 
phase systems of the factorized SO($n$) CG coefficients ($3j$-symbols) 
the last factors coincide with the usual CG coefficients ($3j$-symbols) 
of angular momentum theory \cite{JB77,VMK88}.

\section{Semicanonical bases and coupling coefficients of SO($n$)}

Further, going to the semicanonical basis of the symmetric 
(class-one) irrep $l$ for the chain SO$(n)\!\supset $ SO$(n^{\prime 
})\times $SO$(n^{\prime \prime })\!\supset $SO$(n^{\prime }\!-\!1)
\times $SO$(n^{\prime \prime }-1)\!\supset \cdot \cdot \cdot $, we 
introduce special matrix elements $D_{l^{\prime }M^{\prime },l^{\prime 
\prime }M^{\prime \prime };0}^{n:n^{\prime },n^{\prime \prime };l}(g)$ 
depending only on the rotation angles $\theta _{n^{\prime }-1}^{\prime },
...,\theta _1^{\prime }$ and $\theta _{n^{\prime \prime }-1}^{\prime 
\prime },...\theta _1^{\prime \prime }$ of subgroups SO($n^{\prime }$) 
and SO($n^{\prime \prime }$) and the rotation angle $\theta _c$ in 
$(x_n,x_{n^{\prime }})$ plane, with the second matrix index taken to be 
zero as the $(n-2)$-tuple (0,...,0) for scalar of SO($n-1$). These matrix 
elements may be factorized as follows: 
\begin{equation}
D_{l^{\prime }M^{\prime },l^{\prime \prime }M^{\prime \prime
};0}^{n:n^{\prime },n^{\prime \prime };l}(g)=t_{(n^{\prime })l^{\prime
}0,(n^{\prime \prime })l^{\prime \prime }0;\,(n-1)0}^{(n)\,l}(\theta
_c)D_{M^{\prime }0}^{n^{\prime },l^{\prime }}(g^{\prime })D_{M^{\prime
\prime }0}^{n^{\prime \prime },l^{\prime \prime }}(g^{\prime \prime }).
\label{dmeln}
\end{equation}

Instead of the wavefunction $\Psi _{k,l^{\prime \prime },l^{\prime
}}^{b,a}(\theta _c)=\Psi _{(l-l^{\prime }-l^{\prime \prime })/2,
l^{\prime \prime },l^{\prime }}^{l^{\prime \prime }+n^{\prime \prime }/2
-1,l^{\prime }+n^{\prime }/2-1}(\theta _c)$ (of the type 2c, see (2.6) 
of \cite{IPSW01}) of the tree technique after renormalization with factor 
\[ \left[ \frac{\Gamma (n^{\prime }/2)\Gamma (n^{\prime \prime
}/2)\,d_{l^{\prime }}^{(n^{\prime })}d_{l^{\prime \prime }}^{(n^{\prime
\prime })}}{2\,\Gamma (n/2)\,d_l^{(n)}}\right] ^{1/2} \]
for the integration over the group volume ($0\leq \theta _c\leq \pi /2$)
with measure 
\[ 2{\rm B}^{-1}(n^{\prime }/2,n^{\prime \prime }/2)\sin ^{n^{\prime 
\prime }-1}\theta _c\cos ^{n^{\prime }-1}\theta _c\d 
\theta _c, \]
we obtain special matrix elements of the SO($n$) irreducible 
representation $l$ in terms of the Jacobi polynomials 
\bea
& {t_{(n^{\prime })l^{\prime }0,(n^{\prime \prime })l^{\prime \prime
}0;\,(n-1)0}^{(n)\,l}(\theta _c)=(-1)^{\varphi _{n^{\prime }
n^{\prime \prime }}}\dsty \left[ \frac{d_{l^{\prime }}^{(n^{\prime })}
d_{l^{\prime \prime }}^{(n^{\prime \prime })}\Gamma (n/2)}{d_l^{(n)}
\Gamma (n^{\prime }/2)\Gamma (n^{\prime \prime }/2)\,}\right] ^{1/2}
{\cal N}_{l:l^{\prime },l^{\prime \prime }}^{(n:n^{\prime },
n^{\prime \prime })}} &  \nonumber \\
& {\times \sin ^{l^{\prime \prime }}\theta _c\cos ^{l^{\prime }}
\theta _cP_{(l-l^{\prime }-l^{\prime \prime })/2}^{(l^{\prime \prime }
+n^{\prime \prime }/2-1,l^{\prime }+n^{\prime }/2-1)}(\cos 2\theta _c),} &  
\label{tmeln}
\eea
where the left-hand SO$(n^{\prime })\times $SO$(n^{\prime \prime }$) 
labels are $l^{\prime },l^{\prime \prime }$ ($n^{\prime }
+n^{\prime \prime }=n$), the left-hand SO$(n^{\prime }-1)\times $SO%
$(n^{\prime \prime}-1)$ and right-hand SO($n-1$) labels are 0 for 
rotation with angle $\theta _c$ in $(x_n,x_{n^{\prime }})$ plane. Here 
phase $\varphi _{n^{\prime }n^{\prime \prime }}=0$, unless 
$n^{\prime \prime }=2$, or $n^{\prime }=2$, when the left-hand side 
should be replaced, respectively, by $t_{(n-2)l^{\prime }0,(2)
m^{\prime \prime };\;(n-1)0}^{(n)\;l}(\theta _c)$ with 
$l^{\prime \prime }=|m^{\prime \prime }|$, or by 
$t_{(2)m^{\prime },(n-2)l^{\prime \prime }0;\;(n-1)0}^{(n)\;l}(
\theta _c)$ with $l^{\prime }=|m^{\prime }|$ and
\[ \varphi _{n^{\prime }n^{\prime \prime }}=\tfrac 12[\delta _{n^{\prime
\prime }2}(l^{\prime \prime }-m^{\prime \prime })+\delta _{n^{\prime
}2}(l^{\prime }-m^{\prime })] \]
on the right-hand side and normalization factor 
\begin{equation}
{\cal N}_{l:l^{\prime },l^{\prime \prime }}^{(n:n^{\prime },n^{\prime
\prime })}=\left[ \frac{(l+n/2-1)\left( (l-l^{\prime }-l^{\prime \prime
})/2\right) !\,\Gamma \left( (l+l^{\prime }+l^{\prime \prime }+n
-2)/2\right) }{\Gamma \left( (l-l^{\prime }+l^{\prime \prime }
+n^{\prime \prime })/2\right) \Gamma \left( (l+l^{\prime }
-l^{\prime \prime }+n^{\prime })/2\right) }\right] ^{1/2}.  \label{ncn}
\end{equation}

The $3j$-symbols for the chain SO$(n)\!\supset $SO$(n^{\prime })\times $%
SO$(n^{\prime \prime })\!\supset $SO$(n^{\prime }-1)\times $SO$(n^{\prime
\prime }-1)\!\supset \cdot \cdot \cdot $, labelled by the sets 
$M_i=(l_i^{\prime },N_i^{\prime };l_i^{\prime \prime },
N_i^{\prime \prime })$ may be factorized as follows: 
\bea
& {\left( \begin{array}{ccc}
l_1 & l_2 & l_3 \\ 
M_1 & M_2 & M_3
\end{array} \right) _{\!n}=\left( \begin{array}{ccc}
l_1 & l_2 & l_3 \\ 
l_1^{\prime },l_1^{\prime \prime } & l_2^{\prime },l_2^{\prime
\prime } & l_3^{\prime },l_3^{\prime \prime }
\end{array} \right) _{\!(n:n^{\prime }n^{\prime \prime })}} &  \nonumber \\
& {\times \left( \begin{array}{ccc}
l_1^{\prime } & l_2^{\prime } & l_3^{\prime } \\ 
N_1^{\prime } & N_2^{\prime } & N_3^{\prime }
\end{array} \right) _{\!n^{\prime }}\left( \begin{array}{ccc}
l_1^{\prime \prime } & l_2^{\prime \prime } & 
l_3^{\prime \prime } \\ N_1^{\prime \prime } & N_2^{\prime \prime 
} & N_3^{\prime \prime } 
\end{array} \right) _{\!n^{\prime \prime }}.} &  \label{c3jnc}
\eea
Now the SO$(n)\!\supset $SO$(n^{\prime })\times $SO$(n^{\prime \prime })$
isofactor of $3j$-symbol is expressed as follows: 
\bea
& {\left( \begin{array}{ccc}
l_1 & l_2 & l_3 \\ 
l_1^{\prime },l_1^{\prime \prime } & l_2^{\prime },l_2^{\prime
\prime } & l_3^{\prime },l_3^{\prime \prime }
\end{array} \right) _{\!(n:n^{\prime }n^{\prime \prime })}=\left( 
\begin{array}{ccc}
l_1 & l_2 & l_3 \\ 0 & 0 & 0
\end{array} \right) _{\!n}^{-1}\left( \begin{array}{ccc}
l_1^{\prime } & l_2^{\prime } & l_3^{\prime } \\ 0 & 0 & 0
\end{array} \right) _{\!n^{\prime }}} & \nonumber \\
& {\times \left( \begin{array}{ccc}
l_1^{\prime \prime } & l_2^{\prime \prime } & 
l_3^{\prime \prime } \\ 0 & 0 & 0
\end{array}
\right) _{\!n^{\prime \prime }}\dsty \prod_{a=1}^{3}{\cal N}_{l_a;
l_a^{\prime },l_a^{\prime \prime }}^{(n:n^{\prime },n^{\prime \prime })}
\left[ \frac{d_{l_a^{\prime }}^{(n^{\prime })}d_{l_a^{\prime \prime }}^{%
(n^{\prime \prime })}}{d_{l_a}^{(n)}}\right] ^{1/2}} &  \nonumber \\
& {\times {\rm B}^{1/2}(n^{\prime }/2,n^{\prime \prime }/2)\;
\widetilde{\cal I}\left[ \begin{array}{cccc}
\alpha _0,\beta _0 & \alpha _1,\beta _1 & 
\alpha _2,\beta _2 & \alpha _3,\beta _3 \\ 
& k_1 & k_2 & k_3
\end{array} \right] ,} &  \label{isfnc}
\eea
in terms of auxiliary $3j$-symbols (\ref{isf0}) of the canonical bases
[turning into phase factors of the type \ref{isf0o} for $n^{\prime }=2$ 
or $n^{\prime \prime }=2$], normalization factors (\ref{ncn}) and the 
integrals involving the triplets of Jacobi polynomials 
(\ref{iJp})--(\ref{iJpd}), with parameters 
\begin{eqnarray*}
& {k_i=\tfrac 12(l_i-l_i^{\prime }-l_i^{\prime \prime }),\quad 
\alpha _i=l_i^{\prime \prime }+n^{\prime \prime }/2-1,\quad 
\beta _i=l_i^{\prime }+n^{\prime }/2-1,} & \\
& {\alpha _0=n^{\prime \prime }/2-1,\quad \beta _0=n^{\prime }/2-1} &
\end{eqnarray*}
and 
\begin{eqnarray*}
& {p_i^{\prime }=\tfrac 12(l_j^{\prime }+l_k^{\prime }
-l_i^{\prime }),\quad p_i^{\prime \prime }=\tfrac 12(l_j^{\prime 
\prime }+l_k^{\prime \prime }-l_i^{\prime \prime }),} & \\ 
& {p_i=\tfrac 12(l_j+l_k-l_i)\quad \quad (i,j,k=1,2,3).} &
\end{eqnarray*}
The number of terms in expansion (\ref{iJpd}) of the integrals involving 
triplets of Jacobi polynomials never exceeds 
\begin{subequations} \begin{equation}
B_i=\min \left( \tfrac{1}{6}(p_i+1)_3,(p_i^{\prime \prime }+1)(k_j+1)
(k_{k}+1),\tfrac 12(p_i^{\prime \prime }+1)(p_i+1)_2)\right)   
\label{ntnc}
\end{equation}
and decreases in the intermediate region (e.g., when 
$p_i^{\prime \prime }<p_i+1$), described by the volume of the obliquely 
truncated rectangular parallelepiped of 
$(p_i^{\prime \prime }+1)\times (k_j+1)\times (k_{k}+1)$ size. 

In particular, in the case of $n^{\prime \prime }=2$ parameters 
$l_1^{\prime \prime },l_2^{\prime \prime },l_3^{\prime \prime }$ in 
$3j$-symbol (\ref{isfnc}) should be replaced by $m_1^{\prime \prime }
=\pm l_1^{\prime \prime },m_2^{\prime \prime }=\pm l_2^{\prime 
\prime },m_3^{\prime \prime }=\pm l_3^{\prime \prime }$ so that 
$m_1^{\prime \prime }+m_2^{\prime \prime }+m_3^{\prime \prime }=0$.
Since at least one parameter $p_{i^{\prime }}^{\prime \prime }=0$, the 
number of terms in the $i^{\prime }$th double sum version of 
(\ref{iJpd}) [related to (\ref{iJra}) and to the Kamp\'{e} de 
F\'{e}riet \cite{K-F21,AK-F26} function $F_{2:1}^{2:2}$] does not exceed 
\begin{equation}
\widetilde{B}_{i^{\prime }}=\min \left( \tfrac 12(p_{i^{\prime
}}+1)_2,(k_{j^{\prime }}+1)(k_{k^{\prime }}+1)\right) ,  \label{ntsc}
\end{equation} \end{subequations}
although the $i$th version of (\ref{iJpd}) or (\ref{iJrb}) may 
be more preferable for small values of $p_i$ for which $B_i<\widetilde{B}%
_{i^{\prime }}$.

We may also express the isofactors of the CG coefficients for restriction 
SO$(n)\!\supset $SO$(n^{\prime })\times $SO$(n^{\prime \prime })$ in 
terms of the isofactors of $3j$-symbols, 
\bea
& \left[ \begin{array}{ccc}
l_1 & l_2 & l_3 \\ 
\!l_1^{\prime },l_1^{\prime \prime }\! & \!l_2^{\prime },l_2^{\prime
\prime }\! & \!l_3^{\prime },l_3^{\prime \prime }\!
\end{array}
\right] _{(n:n^{\prime }n^{\prime \prime })} & \nonumber \\
& {=(-1)^{\varphi }\dsty \left[ %
\frac{d_{l_3}^{(n)}}{\!d_{l_3^{\prime }}^{(n^{\prime })}d_{l_3^{%
\prime \prime }}^{(n^{\prime \prime })\!}}\right] ^{1/2}\!\!\!\left( 
\begin{array}{ccc}
l_1 & l_2 & l_3 \\ 
\!\!l_1^{\prime },l_1^{\prime \prime }\! & \!l_2^{\prime },l_2^{\prime
\prime }\! & \!l_3^{\prime },l_3^{\prime \prime }\!\!
\end{array} 
\right) _{\!\!(n:n^{\prime }n^{\prime \prime })},} &  \label{isfncg}
\eea
with the phase $\varphi =0$ (since $l_3-l_3^{\prime }-l_3^{\prime
\prime }$ is even), when $\psi _n,\psi _{n^{\prime }},\psi _{n^{\prime
\prime }}$ are taken to be equal to $J,J^{\prime },J^{\prime \prime }$,
respectively, in all the auxiliary $3j$-symbols (\ref{isf0}), in contrast
to 
\[ \varphi =m_3^{\prime \prime }\delta _{n^{\prime \prime }2}+m_3^{\prime
}\delta _{n^{\prime }2}+l_3^{\prime \prime }\delta _{n^{\prime \prime
}3}+l_3^{\prime }\delta _{n^{\prime }3}, \]
appearing when $\psi _n$ is taken to be zero for $n\geq 4$. Again we 
need to replace, respectively, for $n^{\prime \prime }=2$ parameters 
$l_1^{\prime \prime },l_2^{\prime \prime },l_3^{\prime \prime }$ 
on the left-hand side by $m_1^{\prime \prime },m_2^{\prime \prime },
m_3^{\prime \prime }$ so that $l_1^{\prime \prime }=|m_1^{\prime 
\prime }|,l_2^{\prime \prime }=|m_2^{\prime \prime }|,
l_3^{\prime \prime }=|m_3^{\prime \prime }|$ (with 
$m_1^{\prime \prime }+m_2^{\prime \prime }=m_3^{\prime \prime }$) 
and in the right-hand side by $m_1^{\prime \prime },m_2^{\prime \prime
},-m_3^{\prime \prime }$, as well as for $n^{\prime }=2$ parameters 
$l_1^{\prime },l_2^{\prime },l_3^{\prime }$ on the left-hand side by 
$m_1^{\prime },m_2^{\prime },m_3^{\prime }$ so that $l_1^{\prime
}=|m_1^{\prime }|,l_2^{\prime }=|m_2^{\prime }|,l_3^{\prime
}=|m_3^{\prime }|$ ($m_1^{\prime }+m_2^{\prime }=m_3^{\prime }$) 
and on the right-hand side by 
$m_1^{\prime },m_2^{\prime },-m_3^{\prime }$.

Regarding the different triple-sum versions (\ref{iGpa})--(\ref{iGpd}) 
of integrals involving triplets of Gegenbauer and Jacobi polynomials and 
comparing expressions (\ref{isfnc}) and (\ref{isfc}), we derive the 
following duplication relation between the generic SO$(n)\!\supset $SO$%
(n-1)$ and special SO$(2n+2)\!\supset $SO$(n-1)\times $SO$(n-1)$ 
isofactors of the $3j$-symbols: 
\bea
& {\left( \begin{array}{ccc}
2l_1 & 2l_2 & 2l_3 \\ 
\!l_1^{\prime },l_1^{\prime } & \!l_2^{\prime },l_2^{\prime }\! & 
l_3^{\prime },l_3^{\prime }\!
\end{array} \right) _{\!\!(2n-2:n-1,n-1)}=\dsty \prod_{a=1}^{3}
\left[ \frac{d_{l_a}^{(n)}d_{l_a^{\prime }}^{(n-1)}}{d_{2l_a}^{(2n-2)}}%
\right] ^{1/2}} &  \nonumber \\
& {\times \left( \begin{array}{ccc}
\!2l_1 & \!2l_2\! & 2l_3\! \\ 0 & 0 & 0
\end{array} \right) _{\!\!2n-2}^{-1} \left( \begin{array}{ccc}
l_1 & l_2 & l_3 \\ 0 & 0 & 0
\end{array} \right) _{\!n}} &  \nonumber \\
& {\times \left( \begin{array}{ccc}
l_1^{\prime } & l_2^{\prime } & l_3^{\prime } \\ 
0 & 0 & 0
\end{array}
\right) _{\!n-1}\left( \begin{array}{ccc}
l_1 & l_2 & l_3 \\ 
l_1^{\prime } & l_2^{\prime } & l_3^{\prime }
\end{array} \right) _{\!\!(n:n-1)},} &  \label{isfd}
\eea
with auxiliary $3j$-symbols (\ref{isf0}) of the canonical bases and the 
irrep dimensions appearing.

\section{Basis states and coupling coefficients of the class-two \\
representations of U($n$)}

Mixed tensor irreducible representations $[p+q,q^{n-2},0]\equiv [p,%
\dot{0},-q]$ of U($n$) containing scalar irrep $[q^{n-1}]\equiv 
[\dot{0}]$ of subgroup U($n-1$) (with repeating zeros denoted by 
$\dot{0}$) are called class-two irreps \cite{K74,NAl75}; their canonical 
basis states for the chain U$(n)\!\supset $U$(n-1)\times $U$(1)\!\supset 
\cdot \cdot \cdot \supset $U$(2)\times $U$(1)\!\supset $U(1) are labelled 
by the set 
\begin{eqnarray*}
& {Q_{(n)}=(p_{(n-1)},q_{(n-1)};Q_{(n-1)}^{\prime })} & \\
& {=(p_{(n-1)},q_{(n-1)};p_{(n-2)},q_{(n-2)};...,p_{(2)},q_{(2)};
p_{(1)}),} &
\end{eqnarray*}
where 
\[ p=p_{(n)}\geq p_{(n-1)}\geq ...\geq p_{(2)}\geq 0\quad {\rm and}\quad 
q=q_{(n)}\geq q_{(n-1)}\geq ...\geq q_{(2)}\geq 0 \]
are integers, with $p_{(2)}\geq p_{(1)}\geq -q_{(2)}$ in addition, and
parameters 
\[ M_{(1)}=p_{(1)},\; M_{(2)}=p_{(2)}-q_{(2)}-p_{(1)},...,%
\;M_{(r)}=p_{(r)}-q_{(r)}-p_{(r-1)}+q_{(r-1)} \]
which correspond to irreps of subgroups U$(1)$, beginning from the last
one.

The dimension of representation space is 
\begin{equation}
d_{[p,\dot{0},-q]}^{(n)}=\frac{(p+q+n-1)(p+1)_{n-2}(q+1)_{n-2}}{(n-1)!
(n-2)!}.  \label{dimu}
\end{equation}

Special matrix elements $D_{Q_{(n)};0}^{n[p,\dot{0},-q]}(g)$ of U($n$) 
irrep $[p,\dot{0},-q]$ with zero as the second index for the scalar of 
subgroup U($n-1$) depend only on the rotation angles $\varphi _n,
\varphi _{n-1},...,\varphi _2,\varphi _1$, where 0$\leq \varphi _i
\leq 2\pi $ corresponds to the $i$th diagonal subgroup U(1) 
($i=1,2,...,n$), and $\theta _n,\theta _{n-1},...,\theta _3,
\theta _2$, where $0\leq \theta _r\leq \pi /2$, corresponds to the 
transformation 
\[ \left| \begin{array}{cc}
\cos \theta _r & \i \sin \theta _r \\ 
\i \sin \theta _r & \cos \theta _r
\end{array} \right| \]
in the plane of $(r-1)$st and $r$th coordinates ($r=2,3,...,n$) and may 
be factorized as follows: 
\begin{equation}
D_{Q_{(n)};0}^{n[p,\dot{0},-q]}(g)=\e ^{\i M_n\varphi _n}D_{[p^{\prime 
},\dot{0},-q^{\prime }]0;0}^{n[p,\dot{0},-q]}(\theta _n)D_{Q_{(n-1)};
0}^{n-1[p^{\prime },\dot{0},-q^{\prime }]}(g^{\prime })  \label{dmelu}
\end{equation}
with appropriate normalization in the case of integration over the group
volume with measure 
\[ \frac{(n-1)!}{2\pi ^{n}}\prod_{r=2}^{n}\sin ^{2r-3}\theta _r\cos \theta
_r\d \theta _r\prod_{i=1}^{n}\d \varphi _i. \]
Here $D_{Q_{(n-1)};0}^{n-1[p^{\prime },\dot{0},-q^{\prime }]}
(g^{\prime })$ are the matrix elements of U($n-1$) irrep $[p^{\prime },
\dot{0},-q^{\prime }]=[p_{(n-1)},\dot{0},-q_{(n-1)}]$ (with parameters 
obtained after omitting $\varphi _n$ and $\theta _n$). Special matrix 
elements of U($r$) irreducible representation $[p,\dot{0},-q]$ with the 
U($r-1$) irrep labels $[p^{\prime },\dot{0},-q^{\prime }]$ and 0 and 
SU($r-2$) irrep label 0 for rotation with angle $\theta _r$ in the 
$(x_r,x_{r-1})$ plane are written in terms of the $D$-matrices of SU(2) 
as follows: 
\bea
& {D_{[p^{\prime },\dot{0},-q^{\prime }]0;0}^{r[p,\dot{0},-q]}
(\theta _r)=\left[ (p+q+r-1)\,d_{[p^{\prime },\dot{0},
-q^{\prime }]}^{(r-1)}\right] ^{1/2}\left[ (r-1)\,d_{[p,\dot{0},-q]}^{(r)}
\right] ^{-1/2}} &  \nonumber \\
& {\times (\i \sin \theta _r)^{-r+2}P_{p^{\prime }+(q-p+r-2)/2,
-(p-q+r-2)/2-q^{\prime }}^{(p+q+r-2)/2}(\cos 2\theta _r)} &  \label{tmelud}
\eea
and further, taking into account the identity $P_{m,n}^{l}(x)=
P_{-n,-m}^{l}(x)$, in terms of the Jacobi polynomials 
\bea
& {D_{[p^{\prime },\dot{0},-q^{\prime }]0;0}^{r[p,\dot{0},-q]}
(\theta _r)={\cal N}_{ [p^{\prime },\dot{0},-q^{\prime }]}^{r[p,\dot{0},
-q]}\left[ d_{ [p^{\prime },\dot{0},-q^{\prime }]}^{(r-1)}
\left( (r-1)\,d_{ [p,\dot{0},-q]}^{(r)}\right) ^{-1}\right] ^{1/2}} &  
\nonumber \\
& {\times (\i \sin \theta _r)^{p^{\prime }+q^{\prime }}(\cos 
\theta _r)^{|M|}P_{K}^{(L^{\prime }+r-2,|M|)}(\cos 2\theta _r),} &  
\label{tmeluj}
\eea
where 
\[ K=\min (p-p^{\prime },q-q^{\prime }),\quad M=p-q-p^{\prime }
+q^{\prime },\quad L^{\prime }=p^{\prime }+q^{\prime } \]
and 
\begin{equation} 
{\cal N}_{ [p^{\prime },\dot{0},-q^{\prime }]}^{r[p,\dot{0},-q]}
=\dsty \left[ \frac{(p+q+r-1)K!(p+q+r-2-K)!}{\left( |M|+K\right) !
(p+q+r-2-|M|-K)!}\right] ^{1/2}.  \label{ncua}
\end{equation}
Factor $\i ^{p^{\prime }+q^{\prime }}$ (appearing also in \cite{NAl75}, 
but absent in the generic expressions of $D$-matrix elements 
\cite{Kl79,Vi74}), ensures the complex conjugation relation 
\begin{equation}
\overline{D_{ [p^{\prime },\dot{0},-q^{\prime }]0;0}^{r[p,\dot{0},-q]}
(\theta _r)}=(-1)^{p^{\prime }+q^{\prime }}D_{ [q^{\prime },\dot{0},
-p^{\prime }]0;0}^{r[q,\dot{0},-p]}(\theta _r),  \label{ccur}
\end{equation}
in accordance with the SU(2) case and the system of phases of Baird and
Biedenharn \cite{BB64}, which is correlated to the positive signs of 
the Gel'fand--Tsetlin matrix elements \cite{BR77,Kl79,GC50a} 
of the U($n$) generators $E_{r,r-1}$. Alternatively, the states 
$\Psi _{p^{\prime }+q^{\prime },p+q,M}(\theta _r)$, as defined in 
\cite{KMSS92,K74} and related to the hyperspherical harmonics, correspond 
to the Jacobi polynomials with interchanged parameters $\alpha $ and 
$\beta $. Hence the variables are mutually reflected [here and in 
\cite{KMSS92,K74} as $\cos 2\theta _r$ and ($-\cos 2\theta _r$)].

Using the integration over group (cf.\ \cite{Kl79,K74,KlG79}), the
corresponding $3j$-symbols of the class-two irreps for the chain 
U$(n)\!\supset $U$(n-1)\times $U$(1)\!\supset \cdot \cdot \cdot \supset $%
U$(2)\times $ U$(1)\!\supset $U(1) may be factorized as follows: 
\begin{subequations} \bea
& {\sum_{\rho }\left( \begin{array}{ccc}
\![p_1,\dot{0},-q_1] & [p_2,\dot{0},-q_2] & 
[p_3,\dot{0},-q_3] \\ 
Q_{1(n)} & Q_{2(n)} & Q_{3(n)}
\end{array} \right) _{\!n}^{\rho }} &  \nonumber \\
& {\times \left( \begin{array}{ccc}
\![p_1,\dot{0},-q_1] & [p_2,\dot{0},-q_2] & 
[p_3,\dot{0},-q_3] \\ 
\ [\dot{0}] & [\dot{0}] & [\dot{0}]
\end{array}
\right) _{\!n}^{\rho }} &  \nonumber \\
& {=\int\limits_{{\rm U}(n)}\d gD_{Q_{1(n)};0}^{n[p_1,
\dot{0},-q_1]}(g)D_{Q_{2(n)};0}^{n[p_2,\dot{0},-q_2]}(g)
D_{Q_{3(n)};0}^{n[p_3,\dot{0},-q_3]}(g)} &  \label{c3jui} \\
& {=\delta _{p_1+p_2+p_3,q_1+q_2+q_3}(-1)^{p_1^{\prime
}+p_2^{\prime }+p_3^{\prime }+K_1+K_2+K_3}(n-1)^{-1/2}} &  
\nonumber \\
& {\times \prod_{a=1}^{3}{\cal N}_{ [p_a^{\prime },\dot{0},
-q_a^{\prime }]}^{n[p_a,\dot{0},-q_a]}\left[ d_{ [p_a^{\prime },
\dot{0},-q_a^{\prime }]}^{(n-1)}\left( d_{ [p_a,\dot{0},-q_a]}^{(n)}%
\right) ^{-1}\right] ^{1/2}} &  \nonumber \\
& {\times \widetilde{\cal I}\left[ \begin{array}{cccc}
0,n\!-\!2 & \!|M_1|,L_1^{\prime }\!+\!n\!-\!2 & \!|M_2|,L_2^{\prime }\!%
+\!n\!-\!2 & \!|M_3|,L_3^{\prime }\!+\!n\!-\!2 \\ 
& K_1 & K_2 & K_3
\end{array} \right] } &  \nonumber \\
& {\times \sum_{\rho ^{\prime }}\left( \begin{array}{ccc}
\![p_1^{\prime },\dot{0},-q_1^{\prime }] & [p_2^{\prime },
\dot{0},-q_2^{\prime }] & [p_3^{\prime },\dot{0},-q_3^{\prime }] \\ 
Q_{1(n-1)}^{\prime } & Q_{2(n-1)}^{\prime } & Q_{3(n-1)}^{\prime }
\end{array} \right) _{\!n-1}^{\rho ^{\prime }}} &  \nonumber \\
& {\times \left( \begin{array}{ccc}
\![p_1^{\prime },\dot{0},-q_1^{\prime }] & [p_2^{\prime },
\dot{0},-q_2^{\prime }] & [p_3^{\prime },\dot{0},-q_3^{\prime }] \\ 
\ [\dot{0}] & [\dot{0}] & [\dot{0}]
\end{array} \right) _{\!n-1}^{\rho ^{\prime }}.} &  \label{c3juf}
\eea \end{subequations}
Here $\rho $ and $\rho ^{\prime }$ are the multiplicity labels of the 
U$(n)$ and U$(n-1)$ scalars in the decompositions $[p_1,\dot{0},
-q_1]\otimes [p_2,\dot{0},-q_2]\otimes [p_3,\dot{0},-q_3]$ and 
$[p_1^{\prime },\dot{0},-q_1^{\prime }]\otimes [p_2^{\prime },
\dot{0},-q_2^{\prime }]\otimes [p_3^{\prime },\dot{0},
-q_3^{\prime }]$. The integral involving the product of three Jacobi 
polynomials that appeared in (\ref{c3juf}) also corresponds to the 
SO$(2n)\!\supset $SO$(2n-2)\times $SO(2) isofactor of $3j$-symbol 
\[ \left( \begin{array}{ccc}
p_1+q_1 & p_2+q_2 & p_3+q_3 \\ 
p_1^{\prime }+q_1^{\prime },M_1 & p_2^{\prime }+q_2^{\prime },
M_2 & p_3^{\prime }+q_3^{\prime },M_3
\end{array} \right) _{(2n:2n-2,2)}, \]
considered in previous section and may be expressed (after some 
permutation of parameters) as double sum by means of (\ref{iJra}) or 
(\ref{iJrb}). For normalization of the corresponding $3j$-symbols of 
U$(n)\!\supset $U$(n-1)$ we may use square root of 
\bea
& {\sum_{\rho }\left[ \left( \begin{array}{ccc}
\![p_1,\dot{0},-q_1] & [p_2,\dot{0},-q_2] & 
[p_3,\dot{0},-q_3] \\ 
\ [\dot{0}] & [\dot{0}] & [\dot{0}]
\end{array} \right) _{\!n}^{\rho }\right] ^{2}=\delta _{p_1+p_2+p_3,
q_1+q_2+q_3}} &  \nonumber \\
& {\times (-1)^{\min (p_1,q_1)+\min (p_2,q_2)+\min (p_3,q_3)}\dsty \frac{%
(n-1)![(n-2)!]^2}{\prod_{a=1}^{3}\left( \min (p_a,q_a)+1\right) _{\!n-2}}}
 &  \nonumber \\
& {\times \widetilde{\cal I}\left[ \begin{array}{cccc}
0,n\!-\!2 & \!|p_1\!-\!q_1|,n\!-\!2 & \!|p_2\!-\!q_2|,n\!-\!2 & 
\!|p_3\!-\!q_3|,n\!-\!2 \\ 
& \min (p_1,q_1) & \min (p_2,q_2) & \min (p_3,q_3)
\end{array} \right] ,} &  \label{isfu0}
\eea
with non-vanishing extreme $3j$-symbols in the left-hand side for a 
single value of the multiplicity label $\rho $, which is not correlated 
with the canonical \cite{BLCC72,BLL85,LBL88} and other (see 
\cite{Al88,Al95,Al96}) external labeling schemata of the coupling 
coefficients of U($n$). In 
contrast to the particular $3j$-symbols (\ref{isf0}) of SO($n$), equation 
(\ref{isfu0}) is summable only in the multiplicity-free cases. In 
addition to three double-sum versions of (\ref{iJra}) and (\ref{iJrb}),
the integral on the right-hand side of (\ref{isfu0}) may be also 
expressed as three different double-sum series by means of (\ref{iJpd}), 
taking into account the symmetry relation (\ref{iJpt}). Of course, 
(\ref{isfu0}) is always positive as an analogue of the denominator 
function of the SU(3) canonical tensor operators 
\cite{BLCC72,BLL85,LBL88,Al99}.

Taking into account (\ref{ccur}) we may also obtain expression for the
Clebsch--Gordan coefficients of class-two representation of U($n$) 
\bea
& {\sum_{\rho }\left[ \begin{array}{cc}
\![p_1,\dot{0},-q_1] & [p_2,\dot{0},-q_2] \\ 
Q_{1(n)} & Q_{2(n)}
\end{array} \left| \begin{array}{c}
\![p,\dot{0},-q] \\ Q_{(n)}
\end{array} \right. \right] _{\!n}^{\rho }} &  \nonumber \\
& {\times \left[ \begin{array}{cc}
\![p_1,\dot{0},-q_1] & [p_2,\dot{0},-q_2] \\ 
\ [\dot{0}] & [\dot{0}]
\end{array} \left| \begin{array}{c}
\![p,\dot{0},-q] \\ \![\dot{0}]
\end{array} \right. \right] _{\!n}^{\rho }} &  \nonumber \\
& {=d_{ [p,\dot{0},-q]}^{(n)}\dsty \int\limits_{{\rm U}(n)}\d %
gD_{Q_{1(n)};0}^{n[p_1,\dot{0},-q_1]}(g)D_{Q_{2(n)};0}^{n[p_2,
\dot{0},-q_2]}(g)\overline{D_{Q_{(n)};0}^{n[p,\dot{0},-q]}(g)},} &  
\label{cgcuf}
\eea
with the integrals involving the product of three Jacobi polynomials and 
the CG coefficients of U($n-1$) of the same type and some phase and irrep 
dimension factors. Particularly, we obtain the following expression for 
isofactors of special SU(3) Clebsch--Gordan coefficients (which perform 
the coupling of the SU(3)-hyperspherical harmonics): 
\bea
& {\left[ \begin{array}{ccc}
(a^{\prime }b^{\prime }) & (a^{\prime \prime }b^{\prime \prime }) & 
(ab)_0 \\ 
(z^{\prime })i^{\prime } & (z^{\prime \prime })i^{\prime }; & (z)i
\end{array} \right] } &  \nonumber \\
& {=\delta _{a^{\prime }+a^{\prime \prime }-a,b^{\prime }+b^{\prime 
\prime }-b}(-1)^{i^{\prime }+i^{\prime \prime }-i+K_1+K_2-K}\,
\frac 12\dsty \left[ \frac{(2i^{\prime }+1)(2i^{\prime \prime }+1)}{(2i
+1)\,d_{(a^{\prime }b^{\prime })}^{(3)}d_{(a^{\prime \prime }b^{\prime 
\prime })}^{(3)}}\right] ^{1/2}} &  \nonumber \\
& {\times \dsty \left\{ \frac{(-1)^{\min (a^{\prime },b^{\prime })+\min (%
a^{\prime \prime },b^{\prime \prime })+\min (a,b)}}{\left( \min (%
a^{\prime },b^{\prime })+1\right) \left( \min (a^{\prime \prime },
b^{\prime \prime })+1\right) \left( \min (a,b)+1\right) }\right. } & 
\nonumber \\
& {\times \left. \widetilde{\cal I}\left[ \begin{array}{cccc}
0,1 & |a^{\prime }-b^{\prime }|,1 & |a^{\prime \prime }
-b^{\prime \prime }|,1 & |a-b|,1 \\ 
& \min (a^{\prime },b^{\prime }) & \min (a^{\prime \prime },
b^{\prime \prime }) & \min (a,b)
\end{array} \right] \right\} ^{-1/2}} &  \nonumber \\
& {\times {\cal N}_{ [i^{\prime }-z^{\prime },-i^{\prime }-z^{\prime
}]}^{3[a^{\prime },0,-b^{\prime }]}{\cal N}_{ [i^{\prime \prime
}-z^{\prime \prime },-i^{\prime \prime }-z^{\prime \prime }]}^{3[%
a^{\prime \prime },0,-b^{\prime \prime }]}{\cal N}_{ [i-z,-i-z]}^{3%
[a,0,-b]}\,\left[ \begin{array}{ccc}
i^{\prime } & i^{\prime \prime } & i \\ 
z^{\prime } & z^{\prime \prime } & z
\end{array} \right] } & \nonumber \\
& {\times \widetilde{\cal I}\left[ \begin{array}{cccc}
0,1 & |M^{\prime }|,2i^{\prime }+1 & |M^{\prime \prime }|,2i^{\prime 
\prime }+1 & |M|,2i+1 \\ 
& K^{\prime } & K^{\prime \prime } & K
\end{array} \right] .} &  \label{isfu3}
\eea
Here $M=a-b+2z,\,K=\min (a+z-i,b-z-i)$ in the notation of 
\cite{Al88,Al96}, with $(a~b)$ for the mixed tensor irreps, where 
$a=p_{(3)},\,b=q_{(3)}$ and the basis states are labelled by the isospin 
$i=\tfrac 12(p_{(2)}+q_{(2)})$, its projection $i_{z}=p_{(1)}-
\frac 12(p_{(2)}-q_{(2)})$ and the parameter 
$z=\tfrac 13(b-a)-\tfrac 12y=\tfrac 12(q_{(2)}-p_{(2)})$ instead of 
the hypercharge $y=p_{(2)}-q_{(2)}-\frac 23(p_{(3)}-q_{(3)})$.

\section{Weight shift operators of Sp(4) or SO(5)}

Before considering the triple-sum series appeared \cite{AlJ71} in 
the multiplicity-free isoscalar factors of Sp(4), we include some 
information about the basis states of symplectic group Sp(4) [SO(5)]. The 
irreducible representations of Sp(4) will be denoted by $\langle K\Lambda 
\rangle $, where the pairs of parameters $K=I_{\max },\Lambda =J_{\max }$ 
correspond to the maximal values of irreps $I$ and $J$ of the maximal 
subgroup SU(2)$\times $SU(2) (see \cite{AlJ71,AlJ69,H65}) and to the 
irreps of SO(5) with the highest weight $[L_1L_2]=[K+\Lambda ,
K-\Lambda ]$ and the branching rules $L_1\geq L_1^{\prime }\geq L_2\geq 
|L_2^{\prime }|$, where $L_1^{\prime }=I+J$ and $L_2^{\prime }=I-J$. The 
dimension of representation space of $\langle K\Lambda \rangle $ is 
\[ \tfrac 16(2K-2\Lambda +1)(2\Lambda +1)(2K+2)(2K+2\Lambda +3). \]

The infinitesimal operators (generators) of Sp(4) may be expressed as 
follows:
\begin{subequations} \bea
& {H_1=\tfrac 12(E_{11}-E_{22}),\quad F_{+0}=E_{12},\quad 
F_{-0}=E_{21},} & \label{fg1} \\
& {H_2=\tfrac 12(E_{33}-E_{44}),\quad F_{0+}=E_{34},\quad 
F_{0-}=E_{43},} & \label{fg2} \\
& {T_{++}=-E_{14}-E_{32},\quad T_{--}=E_{41}+E_{23},} & \nonumber \\ 
& {T_{+-}=E_{13}-E_{42},\quad T_{-+}=E_{31}-E_{24}} &  \label{fg3}
\eea \end{subequations}
in terms of generators of SU(4) which satisfy the defining relations 
$[E_{ik},E_{lm}]=\delta _{kl}E_{im}-\delta _{im}E_{lk}$. Operators 
(\ref{fg1}) and (\ref{fg2}) are generators of subgroups SU(2). Their 
matrix elements are well known from the angular momentum theory 
\cite{JB77,VMK88,BL81}. Operators $T_{++}^{2\alpha }$, 
$T_{-+}^{2\alpha }$, $T_{+-}^{2\alpha }$ and $T_{--}^{2\alpha }$ form 
the extreme components of the double SU(2) irreducible tensor operator 
of rank $\alpha ,\alpha $. The corresponding matrix elements may be 
expressed using the Wigner--Eckart theorem, e.g.
\bea
& {\dsty \QATOPD\langle | {\langle K\Lambda \rangle }{I^{\prime }
M^{\prime }J^{\prime }N^{\prime }}T_{-+}^{2\alpha }\QATOPD| \rangle 
{\langle K\Lambda \rangle }{IMJN}=C_{M,-\alpha ,M-\alpha }^{I,\alpha ,
I^{\prime }}C_{N,\alpha ,N+\alpha }^{J,\alpha ,J^{\prime }}} & \nonumber \\
& {\times \dsty \frac{[(2I+1)(2J+1)]^{1/2}(J+J^{\prime }-\alpha )!
\nabla (\alpha II^{\prime })\nabla (\alpha JJ^{\prime })}{P(K\Lambda 
I^{\prime }J^{\prime })P(K\Lambda IJ)}} & \nonumber \\
& {\times \dsty \sum_{i,j}\frac{(-1)^{J-I^{\prime }-\alpha +i-j}(2i+1)
P^2(K\Lambda ij)}{(2j+1)!(J+J^{\prime }+\alpha -2j)!\nabla ^2(j-
J^{\prime },I^{\prime },i)\nabla ^2(j-J,I,i)},} & \label{tmatr}
\eea
where $\nabla (abc)$ is defined by (\ref{nabla}) and 
\[ P(K\Lambda IJ)=E(K+J,I,\Lambda )\nabla ^{-1}(K-J,I,\Lambda ), \]
\begin{equation}
E(abc)=\left[ (a-b-c)!(a-b+c+1)!(a+b-c+1)!(a+b+c+2)!\right] ^{1/2}.
\label{eabc}
\end{equation}
The sum over $i$ in asymmetric (with respect of the couples 
$I,I^{\prime }$ and $J,J^{\prime }$) expression (\ref{tmatr}) for the 
reduced matrix elements (cf.\ (7) of \cite{AlJ71}) corresponds to a very 
well-poised $_9F_8(1)$ hypergeometric series \cite{Sl66,GR90}, 
but the attempting to rearrange it to a more suitable form was 
unsuccessful (e.g.\ in contrast with summations performed in 
\cite{Al02b}). Nevertheless, the reduced matrix elements (\ref{tmatr}) 
are summable in the SU(2) stretched cases (with $I^{\prime }=I\pm \alpha $
or $J^{\prime }=J\pm \alpha $), as well as for the symmetric irrep 
$\langle K0\rangle $ of Sp(4); (\ref{tmatr}) is proportional to the CG 
coefficient of SU(2) for $I+J=K+\Lambda $ or $I^{\prime }+J^{\prime }=
K+\Lambda $, but $_3F_2(1)$ type series appearing for $|J-I|=K-\Lambda $ 
or $|J^{\prime }+I^{\prime }|=K-\Lambda $, as well as for the symmetric 
irrep $\langle \Lambda \Lambda \rangle $ in the general case, are not 
alternating.

The general weight lowering operators of Sp(4) introduced in \cite{AlJ71} 
allow us to obtain the arbitrary basis states when acting into the 
highest weight state. Relation
\begin{equation}
\QATOPD| \rangle {\langle K\Lambda \rangle }{K\!-\!\alpha ,K\!-\!\alpha ,
\Lambda \!+\!\alpha ,\Lambda \!+\!\alpha }=\left[ \frac{(2K-2\Lambda -
2\alpha )!}{(2\alpha )!(2K-2\Lambda )!}\right] ^{1/2}T_{-+}^{2\alpha }
\QATOPD| \rangle {\langle K\Lambda \rangle }{KK\Lambda \Lambda } 
\label{wsh0}
\end{equation}
(see (8) of \cite{AlJ71}) may be used as the first step. More general 
basis state labelled by the chain Sp(4)$\supset $SU(2)$\times $SU(2) may 
be obtained using expansion
\bea
& {\dsty \QATOPD| \rangle {\langle K\Lambda \rangle }{IIJJ}=\sum_{\alpha ,
\beta }Q[\langle K\Lambda \rangle IJ,\alpha \beta ]F_{-0}^{K-I-\alpha -
\beta }F_{0-}^{\Lambda -J+\alpha -\beta }} & \nonumber \\
& {\times T_{--}^{2\beta }\dsty \QATOPD| \rangle {\langle K\Lambda 
\rangle }{K\!-\!\alpha ,K\!-\!\alpha ,\Lambda \!+\!\alpha ,\Lambda \!+\!%
\alpha }} & \label{wsh1}
\eea 
(see (9) of \cite{AlJ71}). We express the expansion coefficient 
$Q[\langle K\Lambda \rangle IJ,\alpha \beta ]$ in the following form:
\bea
& {Q[\langle K\Lambda \rangle IJ,\alpha \beta ]=(-1)^{2\beta }\dsty 
\frac{E(K+\Lambda ,I,J)\nabla (J,I,K-\Lambda )}{(2\beta )!(\Lambda +J+
\alpha -\beta +1)!}} & \nonumber \\
& {\times \dsty \left[ \frac{(2I+1)!(2J+1)!(2\alpha )!}{(2K+1)!(2\Lambda %
)!(2K+2\Lambda +2)!(2K-2\Lambda -2\alpha )!}\right] ^{1/2}} & \nonumber \\
& {\times \dsty \sum_u\frac{(2K-2\Lambda -u)![(K-\Lambda +I+J-u+1)!]^{-1}%
}{u!(K-\Lambda -I+J-u)!(2\alpha -u)!(\Lambda -J-\alpha -\beta +u)!},} & 
\label{cwsh1}
\eea 
appearing instead of the corresponding coefficient in (11) of \cite{AlJ71}
(or its alternative version), when the nonstandard CG coefficient of SU(2) 
\[ C_{\alpha +\beta -K-1,\Lambda +\alpha -\beta +1,\Lambda -K+2\alpha }^{%
I,J,K-\Lambda } \]
(cf.\ (10) of \cite{AlJ71}) is expressed by means of (13.1c) of \cite{JB77}. 

The weight shift relation (5.3) of \cite{NAl74b}\bea
& {\dsty \QATOPD| \rangle {\langle K\Lambda \rangle }{IIJJ}=
\frac{\nabla (K-\Lambda ,I,J)(K+\Lambda -I-J)!(K+\Lambda +I-J+1)!}{%
E(K+\Lambda ,I,J)}} & \nonumber \\
& {\times \dsty \left[ \frac{(2\Lambda +1)!(2I+1)!(2J+1)!}{(2K+1)!
(2K-2\Lambda )!}\right] ^{1/2}\sum_j\left[ \frac{(2K+2j+2)!}{(2j+1)
(2\Lambda -2j)!}\right] ^{1/2}} & \nonumber \\
& {\times \dsty \frac{(-1)^{K-\Lambda -I-J+2j}}{(2J-2j)!(K-\Lambda -I-J
+2j)!(K-\Lambda +I-J+2j+1)!}} & \nonumber \\
& {\times F_{-0}^{K-\Lambda -I-J+2j}T_{-+}^{2J-2j}\dsty \QATOPD| \rangle 
{\langle K\Lambda \rangle }{K\!-\!\Lambda +j,K\!-\!\Lambda +j,j,j}} & 
\label{wsh2}
\eea 
also will be useful in the next sections.

\section{Semistretched isoscalar factors of the second kind of Sp(4) or 
SO(5)}

Actually, the expression of \cite{AlJ71,Al02a} for the semistretched 
isoscalar factors of the second kind\footnote{Remind \cite{AlJ71} that 
the semistretched isoscalar factors of the first kind (with the coupled 
and resulting irrep parameters matching condition $K_1-\Lambda _1+K_2-
\Lambda _2=K-\Lambda $) are proportional to 9$j$-coefficients of SU(2).}
(with the coupled and resulting irrep parameters matching condition 
$K_1+K_2=K$) for the basis labelled by the chain Sp(4)$\supset $SU(2)%
$\times $SU(2) has been derived in \cite{AlJ71} using the weight lowering 
operator (\ref{wsh1}), expansion of the irreducible tensor operators, 
together with rearrangement formulas of the very well-poised 
hypergeometric $_6F_5(-1)$ series (cf.\ \cite{Sl66}), and may be 
presented in the following form: 
\bea
& \left[ \begin{array}{ccc}
\langle K_1\Lambda _1\rangle & \langle K_2\Lambda _2\rangle & 
\langle K_1+K_2,\Lambda \rangle \\ 
I_1J_1 & I_2J_2 & I\,J
\end{array} \right] & \nonumber \\
& {=(-1)^{\Lambda _1+\Lambda _2-\Lambda }\left[
(2I_1+1)(2J_1+1)(2I_2+1)(2J_2+1)(2\Lambda +1)\right] ^{1/2}} &
\nonumber \\
& {\times \dsty \left[ \frac{\prod_{a=1}^{2}(2K_a-2\Lambda _a)!(2K_a+1)!
(2K_a+2\Lambda _a+2)!}{(2K_1+2K_2-2\Lambda )!(2K_1+2K_2+1)!
(2K_1+2K_2+2\Lambda +2)!}\right] ^{1/2}} &  \nonumber \\
& {\times \dsty \frac{\nabla (K_1+K_2-\Lambda ,I,J)
\Delta (I_1I_2I)\Delta (J_1J_2J)\Delta (\Lambda _1\Lambda _2\Lambda )}{%
\prod_{a=1}^{2}E(K_a+\Lambda _a,I_a,J_a)\nabla (K_a-\Lambda _a,
I_a,J_a)}} & \nonumber \\
& {\times E(K_1+K_2+\Lambda ,I,J)\; \widetilde{\cal S}\left[ 
\begin{array}{cccc}
\alpha _0,\beta _0 & \alpha _1,\beta _1 & \alpha _2,\beta _2 & 
\alpha _3,\beta _3 \\ 
& k_1 & k_2 & k_3
\end{array} \right] .} & \label{c11j5}
\eea
In (\ref{c11j5}) and further we use the notations (\ref{nabla}), 
(\ref{eabc}) and
\begin{equation}
\Delta (abc)=\left[ \frac{(a+b-c)!(a-b+c)!(b+c-a)!}{(a+b+c+1)!}
\right] ^{1/2}, \label{del}
\end{equation}
and the triple sum $\widetilde{\cal S}[\cdot \cdot \cdot ]$ in 
special parametrization: 
\begin{subequations} \bea
& {\widetilde{\cal S}\left[ 
\begin{array}{cccc}
\alpha _0,\beta _0 & \alpha _1,\beta _1 & \alpha _2,\beta _2 & 
\alpha _3,\beta _3 \\ 
& k_1 & k_2 & k_3
\end{array} \right] \equiv \widetilde{\cal S}
\left[ \begin{array}{cccc}
K_1 & j_1^1 & j_1^2 & j_1^3 \\ 
K_2 & j_2^1 & j_2^2 & j_2^3 \\ 
K_1+K_2 & j^1 & j^2 & j^3
\end{array} \right] } & \nonumber \\
& {=\dsty \sum_{z_1,z_2,z_3}\binom{p_0}{p_0^{\prime }\!-\!z_1\!-\!z_2\!-%
\!z_3}\prod_{a=1}^{3}\frac{(-1)^{z_a}(-k_a\!-\!\alpha _a)_{z_a}
(-k_a\!-\!\beta _a)_{k_a-z_a}}{z_a!(k_a-z_a)!}} & \label{s11pa} \\
& {=\dsty \binom{-(2k_i\!+\!\alpha _i\!+\!\beta _i\!+\!2)}{-(k_i+
\alpha _i+1)}\sum_{z_1,z_2,z_3}(-1)^{p_i-p_i^{\prime \prime }+z_1+z_2
+z_3}\binom{p_i^{\prime \prime }}{p_i\!-\!z_1\!-\!z_2\!-\!z_3}} &  
\nonumber \\
& {\times \dsty \frac{(k_i+1)_{z_i}(k_i+\beta _i+1)_{z_i}}{z_i!(2k_i
+\alpha _i+\beta _i+2)_{z_i}}\prod_{a\neq i}\frac{(-k_a\!-\!\beta _a)_{%
z_a}(k_a\!+\!\alpha _a\!+\!\beta _a\!+\!1)_{k_a-z_a}}{z_a!(k_a-z_a)!}} &  
\label{s11pb}
\eea \end{subequations}
(with $j_a^1,j_a^2,j_a^3$, $a=1,2$, and $j^1,j^2,j^3$ corresponding to 
transposed $I_a,J_a,\Lambda _a$ and $I,J,\Lambda $, respectively).
The arguments of binomial coefficients are the non-negative integers.
Here 11 parameters of the left-hand side of (\ref{c11j5}) or (\ref{s11pa})
(corresponding to the array of the $11j$ coefficient \cite{AlJ71,Al02a} 
of Sp(4)) are replaced by 
\begin{eqnarray*}
& {k_1=I_1+I_2-I,\quad k_2=J_1+J_2-J,\quad k_3=
\Lambda _1+\Lambda _2-\Lambda ;} & \\
& {\alpha _0=-2K_2-1,\quad \alpha _1=-2I_2-1,\quad \alpha _2=-2J_2-1,
\quad \alpha _3=-2\Lambda _2-1;} & \\
& {\beta _0=-2K_1-1,\quad \beta _1=-2I_1-1,\quad \beta _2=-2J_1-1,
\quad \beta _3=-2\Lambda _1-1.} &
\end{eqnarray*}
Although parameters $\alpha _j$ and $\beta _j$ ($j=0,1,2,3$) here are 
negative integers, arguments of binomial coefficients and 12 linear 
combinations
\begin{subequations} \bea
& {p_0^{\prime }=\tfrac 12(\beta _0-\beta _1-\beta _2-\beta _3)-1=
j_1^1+j_1^2+j_1^3-K_1,} & \nonumber \\
& {p_0^{\prime \prime }=\tfrac 12(\alpha _0-\alpha _k-\alpha _i
-\alpha _0)-1=j_2^1+j_2^2+j_2^3-K_2,} & \nonumber \\
& {p_0=p_0^{\prime }+p_0^{\prime \prime }-k_1-k_2-k_3=j^1+j^2+j^3-K_1
-K_2} & \label{shc0}\\
& {p_i^{\prime }=\tfrac 12(\beta _j+\beta _k-\beta _i-\beta _0)=
K_1-j_1^j-j_1^k+j_1^i,} & \nonumber \\
& {p_i^{\prime \prime }=\tfrac 12(\alpha _j+\alpha _k-\alpha _i
-\alpha _0)=K_2-j_2^j-j_2^k+j_2^i,} & \nonumber \\
& {p_i=k_j+k_k-k_i+p_i^{\prime }+p_i^{\prime \prime }=K_1+K_2-j^j
-j^k+j^i} & \label{shci}
\eea \end{subequations}
($i,j,k=1,2,3$) are non-negative integers, responding to the branching 
rules. Actually, expression (\ref{s11pb}) may be written in three 
versions.

In spite of parameters $\alpha _a$, $\beta _a$ ($a=0,1,2,3$) accepting 
the mutually excluding values in the sums $\widetilde{\cal S}[\cdot \cdot 
\cdot ]$ and $\widetilde{\cal I}[\cdot \cdot \cdot ]$, there is the 
one-to-one correspondence of the analytical continuation between series 
(\ref{s11pa}) and (\ref{iJpb}), as well as between series (\ref{s11pb}) 
and (\ref{iJpd}). In order to demonstrate it, the corresponding beta 
functions with parameters accepting all negative (integer or 
half-integer) values in (\ref{iJpb}) and (\ref{iJpd}) should be replaced 
by the binomial coefficients in (\ref{s11pa}) and (\ref{s11pb}), 
respectively. The possible zeros or poles may be disregarded, when the 
functions $\widetilde{\cal S}[\cdot \cdot \cdot ]\binom{-\alpha _0-
\beta _0-2)}{-\alpha _0-1)}^{-1}$ and $\widetilde{\cal I}[\cdot \cdot 
\cdot ]{\rm B}^{-1}(\alpha _0+1,\beta _0+1)$ are considered, observing 
that the ratio of the binomial coefficients $\binom{-a-b-2}{-a-1}
\binom{-c-d-2}{-c-1}^{-1}$ with negative integers $a,b,c,d$ in equation 
(\ref{s11pa})--(\ref{s11pb}) appeared from the ratio of the beta 
functions ${\rm B}(a+1,b+1){\rm B}^{-1}(c+1,d+1)$ with parameters 
$a,b,c,d\geq -\frac 12$ in relation (\ref{iJpb})--(\ref{iJpd}). 

We see that the restrictions of summation parameters are more rich 
in (\ref{s11pa}) and (\ref{s11pb}) as in (\ref{iJpb})--(\ref{iJpd}).
For example, all three summation parameters are restricted by 
$p_0^{\prime }$, or by $p_0^{\prime \prime }$ in (\ref{s11pa}), as well 
as by $p_i$, or by $p_0^{\prime }$ in (\ref{s11pb}), taking into account 
that in this case $z_i\leq j_1^i-j_2^i+j^i$. Otherwise, the interval for 
the linear combination of summation parameters $z_1+z_2+z_3$ is 
restricted by $p_0$ in (\ref{s11pa}), as well as by 
$p_i^{\prime \prime }$ in (\ref{s11pb}). Hence, taking into account the 
symmetries there are five possibilities of the completely summable 
expressions for $\widetilde{\cal S}[\cdot \cdot \cdot ]$ and seven cases 
when they turn into double sums, dissimilar with nine cases, related to 
the stretched $9j$ coefficients \cite{Al00,JB77}.

As it was demonstrated in \cite{AlJ69}, the Sp(4) isofactors are 
invariant (up to a sign) or may be mutually related under elements of the 
substitution group, generated by the hook reflections and the hook 
permutations 
\begin{subequations} \bea
& {\langle K\Lambda \rangle \to \langle -K-2,\Lambda \rangle ,} & 
\label{hi1} \\
& {\langle K\Lambda \rangle \to \langle K,-\Lambda -1\rangle ,} & 
\label{hi2} \\
& {\langle K\Lambda \rangle \to \langle \Lambda -1/2,K+1/2\rangle .} & 
\label{hp}
\eea \end{subequations}
Using the hook permutations $\langle K_1\Lambda _1\rangle \to 
\langle \Lambda _1-1/2,K_1+1/2\rangle $ and $\langle K\Lambda \rangle \to 
\langle \Lambda -1/2,K+1/2\rangle $ and ``mirror'' reflections 
$J_1\to -J_1-1$ and $J\to -J-1$ (cf.\ \cite{JB77}) to (\ref{c11j5}), the 
expression for the non-standard semistretched isofactors of the second 
kind for the chain Sp(4)$\supset $SU(2)$\times $SU(2) (with the coupled 
and resulting irreps matching condition $\Lambda _1+K_2=\Lambda $) may 
be presented in the following form: 
\bea
& {\left[ \begin{array}{ccc}
\langle K_1\Lambda _1\rangle & \langle K_2\Lambda _2\rangle & 
\langle K,\Lambda _1+K_2\rangle \\ 
I_1J_1 & I_2J_2 & I\,J
\end{array} \right] =(-1)^{K_1+K_2-K-I_1-J_1+I+J}} & \nonumber \\
& {\times \left[ (2I_1+1)(2J_1+1)(2I_2+1)(2J_2+1)(2K\!+\!2)(2\Lambda _1)!
(2K_1\!+\!2\Lambda _1\!+\!2)!\right] ^{1/2}} &
\nonumber \\
& {\times \dsty \left[ \frac{(2K_2-2\Lambda _2)!(2K_2+1)!(2K_2+2\Lambda _2
+2)!(2K-2\Lambda _1-2K_2+1)!}{(2K_1-2\Lambda _1+1)!(2\Lambda _1+2K_2)!
(2\Lambda _1+2K_2+2K+2)!}\right] ^{1/2}} &  \nonumber \\
& {\times \dsty \frac{\nabla (K_1-\Lambda _1,I_1,J_1)\Delta (I_1I_2I)
\Delta (J_1J_2J)\Delta (K_1+1/2,\Lambda _2,K+1/2)}{\nabla (K-\Lambda _1
-K_2,I,J)\nabla (K_2-\Lambda _2,I_2,J_2)\prod_{a=1}^{2}E(K_a+\Lambda _a,
I_a,J_a)}} & \nonumber \\
& {\times E(\Lambda _1+K_2+K,I,J)\; \widetilde{\cal S}\left[ 
\begin{array}{cccc}
\alpha _0,\hat{\beta }_0 & \alpha _1,\beta _1 & 
\alpha _2,\hat{\beta }_2 & \alpha _3,\hat{\beta }_3 \\ 
& k_1 & \hat{k}_2 & \hat{k}_3
\end{array} \right] } & \label{c11jn}
\eea
(cf.\ (3.4) of \cite{NAl74b}), where parameters of $\widetilde{\cal S}
[\cdot \cdot \cdot ]$ (expressed by means of (\ref{s11pb}) with $i=1$) 
accept the values
\begin{eqnarray*}
& {k_1=I_1+I_2-I,\quad \hat{k}_2=J-J_1+J_2,\quad \hat{k}_3=K_1+
\Lambda _2-K;} & \\
& {\alpha _0=-2K_2-1,\quad \alpha _1=-2I_2-1,\quad \alpha _2=-2J_2-1,
\quad \alpha _3=-2\Lambda _2-1;} & \\
& {\hat{\beta }_0=-2\Lambda _1,\quad \beta _1=-2I_1-1,\quad \hat{\beta }_2
=2J_1+1,\quad \hat{\beta }_3=-2K_1-2.} &
\end{eqnarray*}
Taking into account that in this case 
\begin{eqnarray*}
& {\hat{p}_1^{\prime }=I_1+J_1+\Lambda _1-K_1, \quad p_1^{\prime \prime }=
K_2-\Lambda _2+I_2-J_2,} & \\
& {\hat{p}_1=I+J+\Lambda _1+K_2-K} & 
\end{eqnarray*}
and summation parameter $z_1$ is restricted by $z_1\leq I_1-I_2+I$ (but 
condition $z_1\leq k_1$ cannot be regarded), all three summation 
parameters are restricted by conditions \[ 
\hat{p}_1-\min (K_a-\Lambda _a+I_a-J_a)\leq z_1+z_2+z_3\leq \hat{p}_1 \]
(where $a=1$, or 2) and are fixed for $\hat{p}_1=0$, or $K_1-\Lambda _1
+I_1-J_1=0$, when for $p_1^{\prime \prime }=0$ equation (\ref{c11jn}) 
turns into the double sum. Otherwise, when $\widetilde{\cal S}[\cdot 
\cdot \cdot ]$ in (\ref{c11jn}) is expressed by means of (\ref{s11pa}), 
all three summation parameters are restricted by the defining condition 
for the arguments of the binomial coefficients 
\[ \binom{K-\Lambda _1-K_2+I-J}{K_1-\Lambda _1+I_1-J_1-z_1-z_2-z_3} \]
and are fixed for $K_1-\Lambda _1+I_1-J_1=0$, or 
$I_2+J_2+\Lambda _2-K_2=0$. Hence, the expressions for (\ref{c11jn})
become simpler when responding to 8 branching rules of 12 possible.

Note that a restricting condition of the type $K_a-\Lambda _a+I_a-J_a$ 
ensures the double sum expressions for isofactors (\ref{c11j5}) or 
(\ref{c11jn}), e.g., in the case of a symmetric irrep $\langle 
\Lambda _a\Lambda _a\rangle$, whereas the second branching rule 
never causes the existence of any single sum expression. In general, 
new expressions for standard or non-standard triple series
$\widetilde{\cal S}[\cdot \cdot \cdot ]$ may be generated only by the 
elements of the Sp(4) and SU(2) substitution groups, which perform some 
permutation between parameters (\ref{shc0})--(\ref{shci}), i.e., the 
restricting conditions are not spoiled. For example, the expression for 
the non-standard semistretched isofactors of the second kind of Sp(4)
with the coupled and resulting irreps matching condition $\Lambda _1-K_2=
\Lambda $ (which correspond to isofactors (\ref{c11jn}) after interchange 
of $\langle K_1,\Lambda _1\rangle I_1,J_1$ and $\langle K,\Lambda _1+
K_2\rangle I,J$) may be derived from (\ref{c11j5}) using the substitutions
\begin{subequations}
\begin{equation}
\langle K_1\Lambda _1\rangle \to \langle -\Lambda _1-3/2,-K_1-3/2
\rangle ,\quad \langle K\Lambda \rangle \to \langle -\Lambda -3/2,-K-3/2
\rangle ,  \label{ast2a}
\end{equation}
which do not spoil the restricting parameters (\ref{shc0}) in 
(\ref{s11pa}) and (\ref{shci}) in (\ref{s11pb}), with the exception of 
$p_3^{\prime }$, $p_3^{\prime \prime }$ and $p_3$. Otherwise, parameters 
$\hat{p}_0^{\prime }=K_1+\Lambda _1-I_1-J_1=p_3^{\prime }$, 
$p_0^{\prime \prime }$ and $\hat{p}_0=K+\Lambda _1-K_2-I-J=p_3$ play the 
role of the restricting parameters of (\ref{s11pa}), when the 
substitutions 
\bea
& {\langle K_1\Lambda _1\rangle \to \langle -\Lambda _1-3/2,K_1+1/2
\rangle ,\quad \langle K\Lambda \rangle \to \langle -\Lambda -3/2,K+1/2
\rangle ,} & \nonumber \\
& {I_1\to -I_1-1,\quad I\to -I-1,\quad J_1\to -J_1-1,\quad J\to -J-1,} & 
\label{ast2b}
\eea 
\end{subequations}
are used for (\ref{c11j5}).

As a consequence of rearrangement (\ref{iJpst}), an expression for 
special triple sum of the type (\ref{c11j5}) with coinciding the first 
two rows of the corresponding array is derived in the following form:
\bea
& {\widetilde{{\bf S}}\left[ \begin{array}{cccc}
K_1 & j_1^1 & j_1^2 & j_1^3 \\ 
K_1 & j_1^1 & j_1^2 & j_1^3 \\ 
2K_1 & j^1 & j^2 & j^3
\end{array} \right] } & \nonumber \\
& {=(-1)^{j_1^1+j_1^2+j_1^3-K_1-j^3}[1+(-1)^{j^1+j^2+j^3-2K_1}]
2^{j^1+j^2-2K_1-1}(2j^3-1)!} &  \nonumber \\
& {\times \dsty \frac{1}{j^3!}\prod_{a=1}^{2}\frac{(2j_1^a+j^a+1)!}{%
(2j^a+1)!(2j_1^a-j^a)!}\sum_{x_1,x_2,z_3}\binom{j_1^3-\frac 12(j^3+
\delta _3)+x_3}{x_3}} &  \nonumber \\
& {\times \dsty \frac{(-1)^{x_3}\left( -j_1^3-(j^3+\delta _3)/2\right) _{%
x_3}}{(-j^3+1/2)_{x_3}}\binom{\frac 12(\delta _1+\delta _2-\delta _3)}{%
K_1+j^3-\sum_{a=1}^3(\frac 12j^a-x_a)}} &  \nonumber \\
& {\times \dsty \prod_{a=1}^{2}\frac{\left( j_1^a+(j^a+\delta _a)/2
+1\right) _{x_a}}{(j^a+3/2)_{x_a}}\binom{j_1^a-\frac 12(j^a
+\delta _a)}{x_a}.} &  \label{s11je2}
\eea
Here $\delta _i=0$ or 1, so that $j_1^i-(j^i+\delta _i)/2$ 
($i=1,2,3 $) are integers.

Furthermore, an expression for the following more special triple-sum 
$\widetilde{\cal S}[\cdot \cdot \cdot ]$ of the type (\ref{c11j5}) or 
(\ref{s11je2}) (with coinciding parameters $j_1^1=j_1^2$, $j_1^3=K_1$ in 
addition)
\bea
& {\widetilde{{\bf S}}\left[ \begin{array}{cccc}
j_1^3 & j_1^1 & j_1^1 & j_1^3 \\ 
j_1^3 & j_1^1 & j_1^1 & j_1^3 \\ 
2j_1^3 & j^1 & j^2 & j^3
\end{array} \right] } & \nonumber \\
& {=\dsty \frac{(2j_1^3-2j_1^1)!\,\Gamma (1/2)\Gamma \left( (j^1+j^2+j^3
+1)/2-j_1^3\right) }{2^{4j_1^3+3}\prod_{a=1}^3 j^a!\,\Gamma 
\left( j_1^3+(j^1+j^2+j^3+3)/2-j^a\right) }} & \nonumber \\
& {\times \dsty \sum_{s}\frac{(2j_1^1+j^1+1)!(2j_1^1+j^2+1)!(2j_1^3+j^3
+1)!}{s!(2j_1^3-2j_1^1-s)!\left( j_1^3+(j^1-j^2-j^3)/2-s\right) !} } & 
\nonumber \\
& {\times \dsty \frac{[1+(-1)^{j^1+j^2+j^3-2j_1^3}](-1)^{2j_1^1+j_1^3
-(j^1+j^2+j^3)/2}}{\left( j_1^3+(j^2-j^3-j^1)/2-s\right) !
\left( 2j_1^1-j_1^3+(j^3-j^1-j^2)/2+s\right) !}} & \nonumber \\
& {\times \dsty \frac{(2j_1^1+1/2)_s\Gamma (2j_1^3+3/2-s)}{\left( 2j_1^1
-j_1^3+(j^1+j^2+j^3)/2+s+1\right) !},} & \label{s11je1}
\eea
was similarly derived as a consequence of rearrangement (\ref{iJpsr}).
Although this sum also corresponds to the balanced (Saalsch\"{u}tzian) 
$_4F_3(1)$ type series \cite{Sl66,GR90}, it is not alternating (since 
it includes even numbers of gamma functions or factorials in numerator 
and denominator) and cannot be associated with the $6j$ coefficients of 
SU(2). Note, that the summable case of (\ref{s11je1}) with 
$j_1^3=j_1^1$ corresponds to (41) of \cite{AlJ71}.

\section{Isofactors for coupling of two symmetric irreps of SO($n$) 
in the canonical basis}

Now, taking into account the complementary group relation, we may 
consider the most general isofactors of the CG coefficients of SO($n$) 
($n\geq 5$) for coupling of the two symmetric irreps in the canonical 
basis. In the phase system with $\psi _n=0$, we obtain the following 
relations (cf.\ \cite{NAl74b,Al83,Al87}) for these isofactors:
\begin{subequations} \bea
& {\left[ \begin{array}{ccc}
\!l_1\! & \!l_2\! & \!L_1,L_2\! \\ 
\!l_1^{\prime }\! & \!l_2^{\prime }\! & \!L_1^{\prime },L_2^{\prime }\!
\end{array} \right] _{\!(n:n-1)}=\left[ \begin{array}{ccc}
\!l_1\!-\!r\! & \!l_2\!-\!r\! & \!L_1\!-\!r,L_2\!-\!r\! \\ 
\!l_1^{\prime }\!-\!r\! & \!l_2^{\prime }\!-\!r\! & 
\!L_1^{\prime }\!-\!r,L_2^{\prime }\!-\!r\!
\end{array} \right] _{\!(n+2r:n+2r-1)}} &  \label{isfnr} \\
& {=\left[ \begin{array}{ccc}
\!\left\langle \frac{2l_1+n-5}{4}\frac{2l_1+n-5}{4}\right\rangle \! & 
\!\left\langle \frac{2l_2+n-5}{4}\frac{2l_1+n-5}{4}\right\rangle \! & 
\!\left\langle \frac{L_1+L_2+n-5}{2}\frac{L_1-L_2}{4}\right\rangle \! \\ 
\frac{2l_1^{\prime }+n-5}{4},\frac{2l_1^{\prime }+n-5}{4} & 
\frac{2l_2^{\prime }+n-5}{4},\frac{2l_2^{\prime }+n-5}{4} & 
\frac{L_1^{\prime }+L_2^{\prime }+n-5}{2},\frac{L_1^{\prime }-
L_2^{\prime }}{2}
\end{array} \right] ,} &  \label{isf4sp}
\eea \end{subequations}
which are also valid for half-integer values of $r$ (as the analytical 
continuation relations).

Particularly, for $r=L_2^{\prime }=L_2$ equation (\ref{isfc}), together 
with (\ref{isfcg}), may be generalized as follows:
\bea
& {\left[ \begin{array}{ccc}
l_1 & l_2 & L_1,L_2 \\ 
l_1^{\prime } & l_2^{\prime } & L_1^{\prime },L_2
\end{array} \right] _{\!(n:n-1)}=(-1)^{(l_1-l_1^{\prime }-\delta _1+l_2
-l_2^{\prime }-\delta _2+L_1-L_1^{\prime }-\delta )/2}} & \nonumber \\
& {\times \widetilde{\cal I}\left[ \begin{array}{cccc}
\!-\frac 12,L_2\!+\!\frac{n-3}{2} & \!\delta _1\!-\!\frac 12,
l_1^{\prime }\!+\!\frac{n-3}{2} & \!\delta _2\!-\!\frac 12,
l_2^{\prime }\!+\!\frac{n-3}{2} & \!\delta \!-\!\frac 12,L_1^{\prime }\!%
+\!\frac{n-3}{2}\! \\ 
& \frac 12(l_1\!-\!l_1^{\prime }\!-\!\delta _1) & \frac 12(l_2\!%
-\!l_2^{\prime }\!-\!\delta _2) & \frac 12(L_1\!-\!L_1^{\prime }\!%
-\!\delta )
\end{array} \right] } & \nonumber \\
& {\times \dsty \left[ \frac{(2L_1+n-2)(L_1^{\prime }-L_2)!(L_1+L_2+n-3)!
\,\Gamma (L_2+n/2-1)}{8(L_1-L_2)!(L_1^{\prime }+L_2+n-4)!\;\Gamma \left( 
L_2+(n-3)/2\right) }\right] ^{1/2}} & \nonumber \\
& {\times \dsty \frac{\left[ \Gamma (1/2)(2l_1^{\prime }+n-3)
(2l_2^{\prime }+n-3)\right] ^{1/2}\widetilde{\nabla }_{n[0,1,2,3]}
(l_1,l_2;L_1,L_2 )}{\widetilde{\cal H}_{l_1:l_1^{\prime },\delta _1}^{(n)}
\widetilde{\cal H}_{l_2:l_2^{\prime },\delta _2}^{(n)}\widetilde{\cal H%
}_{L_1:L_1^{\prime },\delta }^{(n)}\widetilde{\nabla }_{n[0,1,2,3]}
(l_1^{\prime },l_2^{\prime };L_1^{\prime },L_2)}} & \label{isfmi}
\eea
(cf.\ (4.1) of \cite{NAl74b}), where 
\begin{equation}
\widetilde{\cal H}_{l_a:l_a^{\prime },\delta _a}^{(n)}=
\left[ \frac{\Gamma \left( \frac 12(l_a-l_a^{\prime }+\delta _a+1)\right) 
\Gamma \left( \frac 12(l_a+l_a^{\prime }-\delta _a+n-1)\right) }{\left( %
\frac 12(l_a-l_a^{\prime }-\delta _a)\right) !\,\Gamma \left( \frac 12
(l_a+l_a^{\prime }+\delta _a+n)-1\right) }\right] ^{1/2},  \label{hnor}
\end{equation}
$\widetilde{\nabla }_{n[0,1,2,3]}(\cdot \cdot \cdot )$ is defined by 
(\ref{nabls}) and integral $\widetilde{\cal I}[\cdot \cdot \cdot ]$ may 
be expressed by means of (\ref{iJpd}) (with $i$ chosen 3, 2, or 1).

Further, for $L_1^{\prime }=L_1$, the partial hook permutations 
$[L_1,L_2]\to [L_2-1,L_1+1]$ and $[L_1^{\prime },L_2^{\prime }]\to 
[L_2^{\prime }-1,L_1^{\prime }+1]$ (cf.\ \cite{W74,AlJ74}) allow us to 
transform equation (\ref{isfmi}) into
\bea
& {\left[ \begin{array}{ccc}
l_1 & l_2 & L_1,L_2 \\ 
l_1^{\prime } & l_2^{\prime } & L_1,L_2^{\prime }
\end{array} \right] _{\!(n:n-1)}=(-1)^{l_1-l_1^{\prime }+(\delta _1+
\delta _2-\delta )/2}} & \nonumber \\
& {\times \widetilde{\cal I}\left[ \begin{array}{cccc}
\!-\frac 12,L_1\!+\!\frac{n-1}{2} & \!\delta _1\!-\!\frac 12,
l_1^{\prime }\!+\!\frac{n-3}{2} & \!\delta _2\!-\!\frac 12,
l_2^{\prime }\!+\!\frac{n-3}{2} & \!\delta \!-\!\frac 12,L_2^{\prime }\!%
+\!\frac{n-5}{2}\! \\ 
& \frac 12(l_1\!-\!l_1^{\prime }\!-\!\delta _1) & \frac 12(l_2\!%
-\!l_2^{\prime }\!-\!\delta _2) & \frac 12(L_2\!-\!L_2^{\prime }\!%
-\!\delta )
\end{array} \right] } & \nonumber \\
& {\times \dsty \left[ \frac{(2L_2+n-4)(L_1-L_2+1)!(L_1+L_2+n-3)!\;
\Gamma (L_1+n/2)}{8(L_1-L_2^{\prime }+1)!(L_1+L_2^{\prime }+n-4)!\,\Gamma 
\left( L_1+(n-1)/2\right) }\right] ^{1/2}} & \nonumber \\
& {\times \dsty \frac{\left[ \Gamma (1/2)(2l_1^{\prime }+n-3)
(2l_2^{\prime }+n-3)\right] ^{1/2}\widetilde{\nabla }_{n-1[3,7]}
(l_1^{\prime },l_2^{\prime };L_1,L_2^{\prime })}{\widetilde{\cal H}_{l_1:
l_1^{\prime },\delta _1}^{(n)}\widetilde{\cal H}_{l_2:l_2^{\prime },
\delta _2}^{(n)}\widetilde{\cal H}_{L_2:L_2^{\prime },\delta }^{(n-2)}
\widetilde{\nabla }_{n[3,7]}(l_1,l_2;L_1,L_2)}} & \label{isfma}
\eea
(cf.\ (4.4) of \cite{NAl74b}), where 
\bea
& {\widetilde{\nabla }_{n[3,7]}(a,b;e,f)=\left[ \left( \frac 12(a-b+e-f)
\right) !\left( \frac 12 (b-a+e-f)\right) !\right. } & \nonumber \\
& {\times \left. \Gamma \left( \frac 12(a-b+e+f+n)-1\right) \Gamma 
\left( \frac 12(b-a+e+f+n)-1\right) \right] ^{1/2}} & \nonumber \\
& {\times \dsty \left[ \frac{\Gamma \left( \frac 12(a+b+e-f+n)\right) 
\left( \frac 12(a+b+e+f)+n-3\right) !}{\left( \frac 12(a+b-e-f)\right) !\,
\Gamma \left( \frac 12(a+b-e+f+n)-1\right) }\right] ^{1/2}} & 
\label{nablt}
\eea
(cf.\ (\ref{nabl0})), but the triple sum $\widetilde{\cal I}[\cdot \cdot 
\cdot ]$ (with spoiled defining conditions $\alpha _a-\alpha _0\geq 0$ 
and $\beta _a-\beta _0 \geq 0$ of integrals (\ref{iJp})) may be expressed 
in this case only by means of (\ref{iJpd}) with $i=3$ ($p_3=\frac 12(l_1+
l_2-L_1-L_2)\geq 0$ and $p_3^{\prime }=\frac 12(l_1^{\prime }+l_2^{\prime 
}-L_1-L_2^{\prime })\geq 0$ being integers), as well as using less 
convenient expressions (\ref{iJpb})--(\ref{iJpf}). The triple sum 
$\widetilde{\cal I}[\cdot \cdot \cdot ]$ cannot be expressed as a version 
of (\ref{iJpd}) with $i=1$, or 2, since $p_1$, $p_1^{\prime }$, $p_2$ and 
$p_2^{\prime }$ are negative in this case.

The rather complicated motivation of the phase choice in (\ref{isfma}), 
passed over in \cite{NAl74b}, may be avoided, since equivalence of 
(\ref{isfma}) for SO(5) with expression (33) of \cite{AlJ71} may be 
proved after the transformation (by means of our relation (\ref{iJpst})) 
of the triple sum
\[ \widetilde{\cal I}\left[ \begin{array}{cccc}
\!L_1\!+\!2,L_1\!+\!2 & l_1^{\prime }\!+\!1,l_1^{\prime }\!+\!1 & 
l_2^{\prime }\!+\!1,l_2^{\prime }\!+\!1 & L_2^{\prime },L_2^{\prime } \\ 
& l_1-l_1^{\prime } & l_2-l_2^{\prime } & L_2-L_2^{\prime }
\end{array} \right] \]
that may be discerned in (33) of \cite{AlJ71} using the parameters 
$L_1=K+\Lambda =I^{\prime }+J^{\prime }$, $L_2=K-\Lambda $, 
$l_1=2\Lambda _1$, $l_2=2\Lambda _2$, $l_1^{\prime }=2I_1$, 
$l_2^{\prime }=2I_2$ and $L_2^{\prime }=2I^{\prime }-K-\Lambda $. Note 
the phase factor $(-1)^{L_2-L_2^{\prime }}$ that appears after 
interchange of the sets $l_1,l_1^{\prime }$ and $l_2,l_2^{\prime }$ in 
(\ref{isfma}), in accordance with the number of performed 
antisymmetrizations. 

In analogy with (23) of \cite{AlJ71}, we may derive the most general 
isofactors of Sp(4) in the SU(2)$\times $ SU(2) basis for coupling of the 
two symmetric irreps $\langle \Lambda _1\Lambda _1\rangle $ and 
$\langle \Lambda _2\Lambda _2\rangle $ from overlaps 
\begin{equation}
\QATOPD\langle | {\langle \Lambda _1\Lambda _1\rangle }{I_1,I_1,I_1,J-I_2}
\QATOPD\langle | {\langle \Lambda _2\Lambda _2\rangle }{I_2,I-I_1,I_2,
I_2}\;\QATOPD| \rangle {\langle K\Lambda \rangle }{II;JJ}, \label{isfov}
\end{equation}
expanded using the weight lowering operators of Sp(4) 
(\ref{wsh1})--(\ref{cwsh1}) or (\ref{wsh2}). Furthermore, using relation 
(\ref{isf4sp}) we express the most general isofactors of SO($n$) for 
coupling of the two symmetric irreps in the canonical basis:
\bea
& {\left[ \begin{array}{ccc}
l_1 & l_2 & L_1,L_2 \\ 
l_1^{\prime } & l_2^{\prime } & L_1^{\prime },L_2^{\prime }
\end{array} \right] _{\!(n:n-1)}=\widetilde{\nabla }_{n-1[1,6]}
(l_1^{\prime },l_2^{\prime };L_1^{\prime },L_2^{\prime })} & \nonumber \\
& {\times \dsty \left[ \frac{(L_1+L_1^{\prime }+n-3)!(L_1+L_2^{\prime }
+n-4)!(L_2+L_1^{\prime }+n-4)!}{(L_1-L_2)!(L_1+L_2+n-4)!(2L_1+n-3)!}
\right. } & \nonumber \\
& {\times \dsty \left. \frac{(L_1-L_2^{\prime }+1)!(L_1-L_1^{\prime })!
(L_2-L_2^{\prime })!(L_1^{\prime }-L_2)!(l_1-l_1^{\prime })!
(l_2-l_2^{\prime })!}{(L_2+L_2^{\prime }+n-5)!(l_1+l_1^{\prime }+n-3)!
(l_2+l_2^{\prime }+n-3)!}\right] ^{1/2}} & \nonumber \\
& {\times \dsty \sum_{l_1^0,l_2^0,L_2^0}\left[ \frac{(2l_1^{\prime }
+n-3)(2l_2^{\prime }+n-3)(l_1+l_1^0+n-3)!(l_2+l_2^0+n-3)!}{(2l_1^0+n-3)
(2l_2^0+n-3)(l_1-l_1^0)!(l_2-l_2^0)!}\right] ^{1/2}} & \nonumber \\
& {\times \dsty \frac{[(L_1+L_2^0+n-4)!(L_2-L_2^0)!(L_1-L_2^0+1)!]^{1/2}
}{[(L_2+L_2^0+n-5)!]^{1/2}\widetilde{\nabla }_{n-1[1,6]}(l_1^0,l_2^0;L_1,
L_2^0)(l_1^0-l_1^{\prime })!(l_2^0-l_2^{\prime })!}} & \nonumber \\
& {\times \dsty \sum_u\frac{(-1)^{(l_1^{\prime }+l_2^{\prime }-
L_1^{\prime }+L_2^{\prime }+L_1-L_2^0-l_1^0-l_2^0)/2}(2L_2+n-5-u)!}{u!
(L_2-L_2^{\prime }-u)!(L_2-L_2^0-u)!(L_1^{\prime }+L_2+n-4-u)!}} & 
\nonumber \\
& {\times \dsty \frac{\left[ \left( (l_1^{\prime }+l_2^{\prime }+
L_1^{\prime }-L_2^{\prime }+L_1-L_2^0-l_1^0-l_2^0)/2+1\right) !\right] %
^{-1}}{\left( (l_1^{\prime }+l_2^{\prime }-L_1^{\prime }+L_2^{\prime }
+L_1+L_2^0-l_1^0-l_2^0)/2-L_2+u\right) !}} & \nonumber \\
& {\times \left[ \begin{array}{ccc}
l_1 & l_2 & L_1,L_2 \\ 
l_1^0 & l_2^0 & L_1,L_2^0
\end{array} \right] _{\!(n:n-1)},} & \label{isfg1}
\eea
which are expanded in terms of the boundary (seed) isofactors 
(\ref{isfma}).

In the second case we obtained other expression for the most general 
isofactors of SO($n$) for coupling of the two symmetric irreps 
\bea
& {\left[ \begin{array}{ccc}
l_1 & l_2 & L_1,L_2 \\ 
l_1^{\prime } & l_2^{\prime } & L_1^{\prime },L_2^{\prime }
\end{array} \right] _{\!(n:n-1)}=\widetilde{\nabla }_{n-1[1,3]}
(l_1^{\prime },l_2^{\prime };L_1^{\prime },L_2^{\prime })} & \nonumber \\
& {\times \dsty \left[ \frac{(L_1-L_2+1)!(L_1-L_1^{\prime })!(L_2-L_2^{%
\prime })!(L_1+L_2^{\prime }+n-4)!}{(L_1+L_2+n-4)!(2L_2+n-5)!(L_1^{\prime 
}-L_2)!(L_1-L_2^{\prime }+1)!}\right. } & \nonumber \\
& {\times \dsty \left. \frac{(L_2+L_2^{\prime }+n-5)!(L_2+L_1^{\prime }
+n-4)!(l_1-l_1^{\prime })!(l_2-l_2^{\prime })!}{(L_1+L_1^{\prime }+n-3)!
(l_1+l_1^{\prime }+n-3)!(l_2+l_2^{\prime }+n-3)!}\right] ^{1/2}} & 
\nonumber \\
& {\times \dsty \sum_{l_1^0,l_2^0,L_1^0}\left[ \frac{(2l_1^{\prime }+n-3)
(2l_2^{\prime }+n-3)(2l_2^0+n-3)(l_1+l_1^0+n-3)!}{(2l_1^0+n-3)(l_1-l_1^0)!
(l_2+l_2^0+n-3)!(L_1-L_1^0)!}\right] ^{1/2}} & \nonumber \\
& {\times \dsty \frac{[(l_2-l_2^0)!(L_1^0-L_2)!(L_1+L_1^0+n-3)!(L_2+L_1^0
+n-4)!]^{1/2}}{2\widetilde{\nabla }_{n-1[1,3]}(l_1^0,l_2^0;L_1^0,L_2)
(l_1^0-l_1^{\prime })!(L_1^0-L_1^{\prime })!(L_1^0+L_1^{\prime }+n-4)!}} &
\nonumber \\
& {\times \dsty \sum_v\frac{(-1)^{l_1^0-l_1^{\prime }}(l_2+l_2^{\prime }+
n-3+u)!}{v!(l_2-l_2^{\prime }-v)!(l_2^{\prime }-l_2^0+v)!\;\Gamma 
\left( l_2^{\prime }+(n-1)/2+v\right) }} & \nonumber \\
& {\times \dsty \frac{\Gamma \left( (l_2^{\prime }-l_1^{\prime }-
L_1^{\prime }+L_2^{\prime }+L_1^0-L_2+l_1^0+l_2^0+n-3)/2\right) }{\left( 
(l_1^{\prime }-l_2^{\prime }+L_1^{\prime }-L_2^{\prime }-L_1^0+L_2-l_1^0
+l_2^0)/2-v\right) !}} & \nonumber \\
& {\times \left[ \begin{array}{ccc}
l_1 & l_2 & L_1,L_2 \\ 
l_1^0 & l_2^0 & L_1^0,L_2
\end{array} \right] _{\!(n:n-1)},} & \label{isfg2}
\eea
expanded in terms of the boundary (seed) isofactors (\ref{isfmi}). Note, 
that the restrictions for summation parameters $l_1^0$ and $l_2^0$ in 
(\ref{isfg2}) are different and an alternative version of it with 
mutually interchanged parameters $l_1,l_1^{\prime },l_1^0$ and 
$l_2,l_2^{\prime },l_2^0$ (but the same phase factor $(-1)^{l_1^0-
l_1^{\prime }}$) is possible. The total number of summation parameters in 
both expressions (\ref{isfg1}) and (\ref{isfg2}) is six, in contrast 
with seven in expansion (4.2) of \cite{Al87} in terms of the boundary 
isofactors 
\[ \left[ \begin{array}{ccc}
l_1 & l_2 & L_1,L_2 \\ 
l^0 & l^0 & L_2,L_2
\end{array} \right] _{\!(n:n-1)}. \]

Isofactors of SO($n$) for the semistretched coupling (with the coupled 
and resulting irrep parameters matching condition $l_1+l_2=L_1+L_2$) may 
be expressed as the double sums, using relation (\ref{isf4sp}), together 
with (\ref{c11j5}) and $i=2$ version of (\ref{s11pb}). Otherwise, 
isofactors of SO($n$) with the coupled and resulting irrep parameters 
matching condition $l_1-l_2=L_1+L_2$ ($l_1\geq l_2$) may be derived 
using relation (\ref{isf4sp}), together with (\ref{c11jn}) and 
the $i=1$ version of (\ref{s11pb}), after applying the symmetry relation 
(A.22) of \cite{H65} (interchange of $\langle K_1,\Lambda _1\rangle 
I_1,J_1$ and $\langle K,\Lambda _1+K_2\rangle I,J$) to isofactors 
(\ref{c11jn}) of Sp(4).

\section{Concluding remarks}

In this paper, we reconsidered once more the $3j$-symbols and 
Clebsch--Gordan coefficients of the orthogonal SO($n$) and unitary U($n$) 
groups for all three representations corresponding to the 
(ultra)spherical or hyperspherical harmonics of these groups (i.e.\ 
irreps induced \cite{BR77} by the scalar representations of the SO($n-1$) 
and U($n-1$) subgroups, respectively). For the corresponding isoscalar 
factors of the $3j$-symbols and coupling coefficients, the ordinary 
integrations involving triplets of the Gegenbauer and the Jacobi 
polynomials yield the more or less symmetric triple-sum expressions, 
however without the apparent triangle conditions. These conditions are 
visible and efficient only in here directly proved expressions 
(\ref{iJpd}), (\ref{iGpc}) and (\ref{iGpr}), previously derived in 
\cite{NAl74a,Al83,Al02a} after complicated analytical continuation 
procedure of special Sp(4)$\supset $SU(2)$\times $SU(2) isofactors (cf.\ 
\cite{AlJ71,AlJ69}). Actually, only for a fixed integer shift parameter 
$p_i=\frac 12(l_j+l_k-l_i)$ it is evident that the corresponding 
integrals involving triplets of the Gegenbauer and the Jacobi polynomials 
are rational functions of remaining parameters. Practically, the concept 
of the canonical unit tensor operators (cf.\ section 21 of chapter 3 of 
\cite{BL81}) for symmetric irreps of SO($n$) may be formulated only under 
such a condition.

Similarly as special terminating double-hypergeometric series of Kamp\'{e}
de F\'{e}riet-type \cite{Al00,K-F21,AK-F26,LV-J01,V-JPR94} 
correspond to the stretched $9j$ coefficients of SU(2), the definite 
terminating triple-hypergeometric series correspond either to the 
semistretched isofactors of the second kind \cite{AlJ71} of Sp(4), or to 
the isofactors of the symmetric irreps of the orthogonal group SO($n$) 
in the canonical and semicanonical (tree type) bases. Relation 
(2.6a)--(2.6c) of \cite{Al02a} (being significant within the framework 
of Sp(4) isofactors) is a triple-sum generalization of transformation 
formula (9) of \cite{LV-J01} for terminating $F_{1:1,1}^{1:2,2}$ 
Kamp\'{e} de F\'{e}riet series with a fixed single-integer non-positive 
parameter, restricting all summation parameters, although this 
termination condition is hidden in our equation (\ref{s11pa})--%
(\ref{s11pb}). Our auxiliary expression (\ref{iJpf}) for integrals 
corresponds to analytical continuation of the intermediate formula 
(2.6b) of \cite{Al02a} (or (26)--(27) of \cite{AlJ71}) for special 
isofactors of Sp(4) or SO(5). However, relation 
(\ref{iJpb})--(\ref{iJpd}) (important within the framework of SO($n$) 
isofactors) cannot be associated with any transformation formula 
\cite{LV-J01} for terminating $F_{1:1,1}^{1:2,2}$
Kamp\'{e} de F\'{e}riet series with the same (single or double) 
parameters, restricting summation. Note the quite different procedures 
for generating the diversity of expressions for integrals and special 
isofactors of Sp(4) (variation of expressions for the Jacobi polynomials 
and use of the substitution group technique).

Expressions (\ref{s11pa}) and (\ref{s11pb}) corresponding to special 
Sp(4) isofactors are summable or turn into the terminating Kamp\'{e} de 
F\'{e}riet \cite{K-F21,AK-F26,LV-J01} series $F_{2:1}^{2:2}$ for extreme 
basis states of Sp(4)$\supset $SU(2)$\times $ SU(2). Alternatively, in 
accordance with (\ref{iGpco}) and (\ref{iJpd}), the expressions for 
special isofactors of SO($n$) and SU($n$) are summable in the case of the 
stretched couplings of the group representations and turn into the 
terminating Kamp\'{e} de F\'{e}riet series $F_{2:1}^{2:2}$ for the irreps 
of subgroups in a stretched situation, including the generic cases for 
restrictions SO$(n)\!\supset $SO($n-1$), 
SO$(n)\!\supset $SO$(n-2)\times $SO(2) and U$(n)\!\supset $U($n-1$). 
Taking into account the fact that the $F_{2:1}^{2:2}$ type series with 
five independent parameters also appeared as the denominator 
(normalization) functions of the SU(3) and $u_q(3)$ canonical tensor 
operators \cite{BLCC72,BLL85,LBL88} (cf.\ (2.8) and section II of 
\cite{Al99}), the $q$-extension of relation (\ref{iJra})--(\ref{iJrb}) 
from the classical SU($n$) case may be suspected. 

The expressions for special isofactors of SO($n$) in terms of $_4F_3(1)$ 
series were helpful for rearrangement \cite{Al02b} of the fourfold 
\cite{Al87,HJu99} and (corrected) \cite{Al87} triple-sum expressions for 
the recoupling coefficients ($6j$-symbols) of symmetric irreps of SO($n$) 
into the double $F_{1:3}^{1:4}$ type series, with the Regge type symmetry.

\end{document}